\documentclass[twocolumn,pre,groupedaddress,amsmath,amssymb]{revtex4}

\usepackage{amsmath, amssymb, amsfonts, enumerate, mathrsfs}
\usepackage{graphicx}
\usepackage{color}

\begin{document}

\title{Many-body effects for critical Casimir forces}

\author{T. G. Mattos}
\email{tgmattos@is.mpg.de}
\affiliation{Max-Planck-Institut f\"ur Intelligente Systeme, Heisenbergstr. 3,
D-70569 Stuttgart, Germany,}
\affiliation{IV. Institut f\"ur Theoretische Physik, Universit\"at
Stuttgart, Pfaffenwaldring 57, D-70569 Stuttgart, Germany}

\author{L. Harnau}
\affiliation{Max-Planck-Institut f\"ur Intelligente Systeme, Heisenbergstr. 3,
D-70569 Stuttgart, Germany,}
\affiliation{IV. Institut f\"ur Theoretische Physik, Universit\"at
Stuttgart, Pfaffenwaldring 57, D-70569 Stuttgart, Germany}

\author{S. Dietrich}
\affiliation{Max-Planck-Institut f\"ur Intelligente Systeme, Heisenbergstr. 3,
D-70569 Stuttgart, Germany,}
\affiliation{IV. Institut f\"ur Theoretische Physik, Universit\"at
Stuttgart, Pfaffenwaldring 57, D-70569 Stuttgart, Germany}

\newpage

\begin{abstract}
Within mean-field theory we calculate the scaling functions associated
with critical Casimir forces for a system consisting of two spherical
colloids immersed in a binary liquid mixture near its consolute point and facing
a planar, homogeneous substrate. For several geometrical arrangements and boundary conditions
we analyze the normal and the lateral critical Casimir forces acting on one of the two colloids.
We find interesting features such as a change of sign of these forces upon varying either the position of one of the colloids or the temperature. By subtracting the pairwise forces from the total force we
are able to determine the many-body forces acting on one of the colloids. We
have found that the many-body contribution to the total critical Casimir force
is more pronounced for small colloid-colloid and colloid-substrate distances,
as well as for temperatures close to criticality, where the many-body
contribution to the total force can reach up to $25\%$.




\end{abstract}

\maketitle

\newpage


\section{Introduction}


The first step in describing the interaction between many particles is to determine
their pair potential or the forces among a single pair. If the governing physical equations are linear (like for gravity or electrostatics), this approach yields a quantitatively reliable description of the physical system
considered, based on the linear superposition principle. However, if nonlinearities are present, linear superposition of pair potentials is no longer accurate and nonadditivity gives rise to many-body effects. These latter effects can lead, e.g., to a strengthening or weakening of the total force acting on a particle surrounded by more than a single other one, a change of sign of that force, or the appearance of stable or unstable configurations. Many-body effects appear in rather diverse systems such as nuclear matter~\cite{THR1958–1959615}, superconductivity~\cite{PhysRev.124.670}, colloidal
suspensions~\cite{PhysRevLett.92.078301,B926845F}, quantum-electrodynamic Casimir
forces~\cite{PhysRevA.77.030101,PhysRevA.80.022519,PhysRevLett.104.160402,
PhysRevA.83.042516,PhysRevLett.99.080401,PhysRevLett.105.040403}, polymers~\cite{refId0,PhysRevE.83.061801}, nematic colloids~\cite{Eur.Phys.J.E21}, and noble
gases with van der Waals forces acting among them~\cite{axilrod:299,Wells1983429,PhysRevLett.57.230,jakse:8504}. Each of these systems is characterized by a wide range of time and length scales. Integrating out the degrees of freedom associated with small scales (such as the solvent of colloidal solutes or polymers) for fixed configurations of the large particles, generates effective interactions among the latter, which are inherently not pairwise additive. This is the price to be paid for achieving a reduced description of a multicomponent system. Driven by these effective interactions the large particles of the system may exhibit collective behavior of their own (like aggregation or phase separation, see Refs.~\cite{mohry:224902} and~\cite{mohry:224903} and references therein), which can be described much easier if it is governed by pair potentials. In order to be able to judge whether this ansatz is adequate one has to check the relative magnitude of genuine many-body forces.

In this paper we assess the quantitative influence of such many-body effects on critical Casimir forces (CCFs)~\cite{krech:1,brankov:1,1742-6596-161-1-012037}.
These long-ranged forces arise as a consequence of the confinement of the order
parameter fluctuations in a critical fluid~\cite{fisher_degennes}. They have been analyzed paradigmatically by studying  the effective interaction between a \textit{single} colloidal particle and a homogeneous~\cite{PhysRevLett.74.3189,PhysRevB.51.13717,PhysRevLett.81.1885,Schlesener:JStatPhys.2003,eisenriegler:3299,kondrat:204902,nature_letters,epnews,PhysRevE.80.061143} or inhomogeneous~\cite{PhysRevLett.101.208301,EPL:troendle.2009,trondle:074702,MolPhys:troendle.2011,C0SM00635A} container wall as well as between \textit{two isolated} colloidal particles~\cite{PhysRevLett.74.3189,PhysRevB.51.13717,PhysRevLett.81.1885,Schlesener:JStatPhys.2003,eisenriegler:3299} upon approaching the critical point of the solvent. Here we add one sphere to the sphere-wall configuration, which is the simplest possibility to study many-body forces. (The wall mimics a third, very large sphere.)

In order to be able to identify the latter ones one has to resort to a theoretical scheme which allows one to compute the forces between individual pairs and the three-body forces on the same footing. Since these forces are characterized by universal scaling functions, which depend on the various geometrical features of the configuration and on the thermodynamic state, we tackle this task by resorting to field-theoretic mean field theory (MFT), which captures the universal scaling functions as the leading contribution to their systematic expansion in terms of $\epsilon =4-d$ spatial dimensions. Experience with corresponding previous studies for simple geometries tells that this approach does yield the relevant qualitative features of the actual universal scaling functions in $d=3$; if suitably enhanced by renormalization group arguments these results reach a semi-quantitative status. We point out that even within this approximation the numerical implementation of this corresponding scheme poses a severe technical challenge. Thus at present this approach appears to be the only feasible one to explore the role of many-body critical Casimir forces within the full range of their scaling variables.

Accordingly we consider the standard Landau-Ginzburg-Wilson Hamiltonian for critical phenomena of the Ising bulk universality class, which is given by

\begin{equation}\label{LGW-Hamiltonian}
\mathscr{H}[\phi] = \int_V{\rm d}^d\mathbf{r} \left\{ \frac{1}{2}\left(
\nabla\phi \right)^2 + \frac{\tau}{2}\phi^2 + \frac{u}{4!}\phi^4 \right\} ~,
\end{equation}

\noindent
with suitable boundary conditions (BCs). In the case of a binary liquid mixture near its consolute (demixing) point, the order parameter $\phi(\mathbf{r})$ is proportional to the deviation of the local concentration of one of the two species from the critical concentration. $V$ is the volume accessible to the fluid, $\tau$ is proportional to the reduced temperature $t=(T-T_c)/T_c$, and the coupling constant $u>0$ stabilizes the statistical weight $\mathrm{exp}(-\mathscr{H})$ in the two-phase region, i.e., for $t<0$. Close to the bulk critical point $T_c$ the bulk correlation length $\xi_{\pm }$ diverges as $\xi_{\pm}(t\rightarrow 0^{\pm}) = \xi_0^{\pm}\left| t\right|^{-\nu}$, where $\nu\simeq 0.63$ in $d=3$ and $\nu = 1/2$ in $d=4$, i.e., within MFT~\cite{Pelissetto2002549}. The two non-universal amplitudes $\xi_0^{\pm}$ are of molecular size; they form the universal ratio $\xi_0^+/\xi_0^-\approx 1.9$ for $d=3$ and $\xi_0^+/\xi_0^- = \sqrt{2}$ for $d=4$~\cite{phasetransition_privman}. The BCs reflect the generic adsorption preference of the confining surfaces for one of the two components of the mixture. For the critical adsorption fixed point~\cite{phasetransition_diehl}, the BC at each of the confining surfaces is either $\phi=+\infty$ or $\phi=-\infty$, to which we refer as $(+)$ or $(-)$, respectively.

Within MFT the equilibrium order parameter distribution minimizes the Hamiltonian in Eq.~\eqref{LGW-Hamiltonian} for the aforementioned BCs, i.e., $\delta\mathscr{H}[\phi]/\delta\phi = 0$. Far from any boundary the order parameter approaches its constant bulk value $\left< \phi \right> = \pm \mathcal{A}\left| t\right|^{\beta}$ for $t<0$ or $\left< \phi \right> = 0$ for $t>0$. $\mathcal{A}$ is a non-universal bulk amplitude and $\beta = 1/2$ (for $d=4$) is a standard critical exponent. In the following we consider the reduced order parameter $m=u^{1/2}\left< \phi \right>$.

The remainder of this paper is organized as follows. In
Sec.~\ref{section_the_system} we define the system under consideration and the scaling functions for the CCFs as well as the normalization scheme. In Sec.~\ref{section_results} we present the numerical results obtained for the universal scaling functions of the CCFs, from which we extract and analyze the many-body effects. In Sec.~\ref{section_conclusion} we summarize our results and draw some conclusions.


\section{Physical system}\label{section_the_system}

\begin{figure}
\begin{center}
\includegraphics[scale=0.7]{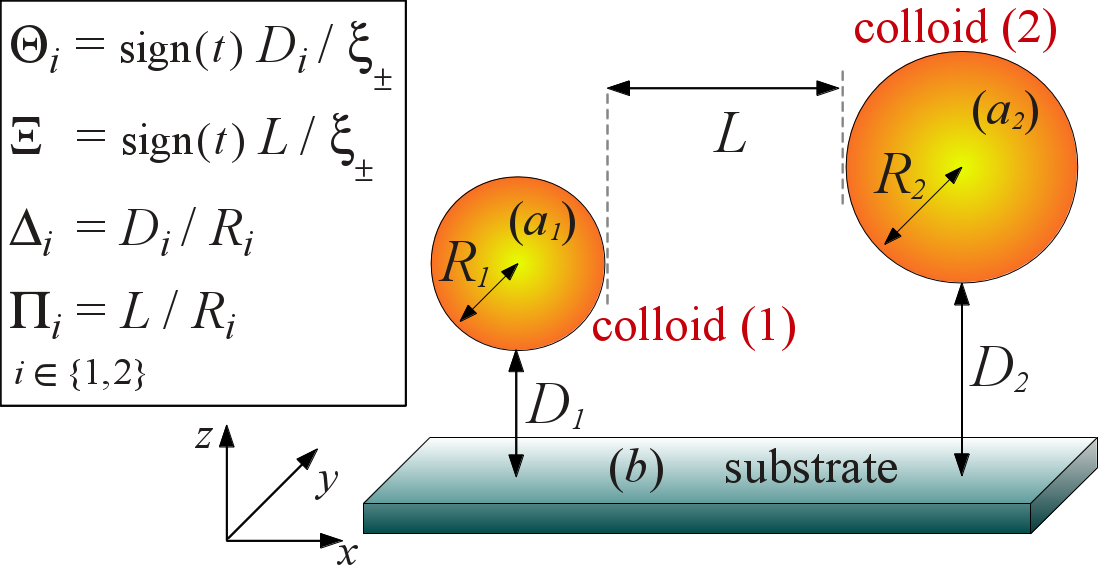}
\caption{
Two spherical colloidal particles of radii $R_1$ and $R_2$ immersed in a near-critical binary liquid mixture (not shown) and close to a homogeneous, planar substrate at $z=0$. The two colloidal particles with BCs $(a_1)$ and $(a_2)$ are located at sphere-surface-to-substrate distances $D_1$ and $D_2$, respectively. The substrate exhibits BC $(b)$. The lateral distance between the centers of the spheres along the $x$-direction is given by $R_1 + L +R_2$, while the centers of both spheres lie in the plane $y=0$. In the case of four spatial dimensions the figure shows a cut of the system, which is invariant along the fourth direction, i.e., the spheres correspond to parallel hypercylinders with one translationally invariant direction, which is $r_4$.}
\label{system_sketch}
\end{center} 
\end{figure}

We study the normal and the lateral CCFs acting on two colloidal particles immersed
in a near-critical binary liquid mixture and close to a homogeneous, planar
substrate. We focus on the critical concentration which implies the absence of a bulk field conjugate to the order parameter [see Eq.~\eqref{LGW-Hamiltonian}]. The surfaces of the colloids and of the substrate are considered to exhibit a strong adsorption preference for one of the two components of the confined liquid leading to $\pm$ BCs. The forces are calculated using the full three-dimensional numerical analysis of the appropriate MFT as given by Eq.~\eqref{LGW-Hamiltonian}. Specifically, we consider two three-dimensional spheres of radii $R_1$ and $R_2$ with BCs $(a_1)$ and ($a_2$), respectively, facing a homogeneous substrate with BC $(b)$ at sphere-surface-to-substrate distances $D_1$ and $D_2$, respectively (see Fig.~\ref{system_sketch}). The coordinate system $(x,y,z)$ is chosen such that the centers of the spheres are located at $(x_1,0,D_1+R_1)$ and $(x_1+R_1+L+R_2,0,D_2+R_2)$ so that the distance between the centers, projected onto the $x$-axis, is given by $R_1 + L +R_2$. The BCs of the whole system are represented by the set $(a_1,a_2,b)$, where $a_1$, $a_2$, and $b$ can be either $+$ or $-$. It is important to point out that we discuss colloidal particles with the shape of a hypercylinder

\begin{widetext}
\begin{equation} \label{eu<3hipercilindros}
\mathcal{C}_d\left( \left\{ \mathcal{R}_s \right\} \right) = \left\{ \mathbf{r} = \left( r_1, r_2,\ldots, r_d  \right) \in  \mathbb{R}^d \left. \right| \sum_{s=1}^{\mathfrak{D}}\left( \dfrac{r_s}{\mathcal{R}_s} \right)^2 \le 1, \quad\mathfrak{D} \le d \right\}~,
\end{equation}
\end{widetext}

\noindent
where $\mathcal{R}_1\le \mathcal{R}_2 \le\ldots\le \mathcal{R}_\mathfrak{D}$ are the semiaxes (or radii) of the hypercylinder and $x=r_1$, $y=r_2$, $z=r_3$. If $\mathcal{R}_1= \mathcal{R}_2 = \ldots = \mathcal{R}_\mathfrak{D}$ and $d=\mathfrak{D}$, the hypercylinder reduces to a hypersphere. The generalization of $d$ to values larger than 3 is introduced for technical reasons because $d_c=4$ is the upper critical dimension for the relevance of the fluctuations of the order parameter. These fluctuations lead to a behavior different from that obtained from the present MFT which (apart from logarithmic corrections~\cite{phasetransition_diehl,0295-5075-74-1-022}) is valid in $d=4$. We consider two hypercylinders in $d=4$ with $\mathcal{R}_1=\mathcal{R}_2=\mathcal{R}_3=R$ and $\mathfrak{D}=3$. The two colloids are taken to be parallel along the fourth dimension with infinitely long hyperaxes in this direction. Considering hypercylinders, which are translationally invariant along the $r_4-$axis, allows us to minimize $\mathscr{H}[\phi]$ numerically using a three dimensional finite element method in order to obtain the spatially inhomogeneous order parameter profile $\phi(x,y,z)$ for the geometries under consideration (see Fig.~\ref{system_sketch}).


In the case of an upper critical demixing point of the binary liquid mixture at
the critical concentration, $t > 0$ corresponds to the disordered (i.e., mixed)
phase of the fluid, whereas $t < 0$ corresponds to the ordered (i.e., phase separated) phase. The meaning of the sign is reversed for a lower critical point. In the following we assume an upper critical point.

The normal CCF $F^{(i,z)}_{(a_1,a_2,b)}(D_1,D_2,R_1,R_2,L,t)$ acting on sphere
$i$ in the presence of sphere $j$ ($\{i,j\} \in  \{1, 2\}$ and $i \neq j$) along the $z$-direction takes the scaling form

\begin{multline} \label{def_scaling_z}
F^{(i,z)}_{(a_1,a_2,b)}(D_1,D_2,R_1,R_2,L,t) =\\
{}= k_BT \frac{R_i}{D_i^{d-\mathfrak{D}+2}}K^{(i,z)}_{(a_1,a_2,b)}(\Theta_i,\Delta_1,\Delta_2,\Pi_1,\Pi_2)\,,
\end{multline}

\noindent
where $\Delta_1=D_1/R_1$, $\Delta_2=D_2/R_2$, $\Pi_1=L/R_1$, $\Pi_2=L/R_2$, and $\Theta_i={\rm sign}(t) D_i/\xi_\pm$ (i.e., $\Theta_i = D_i/\xi_+$ for $t>0$ and $\Theta_i = -D_i/\xi_-$ for $t<0$). Equation \eqref{def_scaling_z} describes the singular contribution to the normal force emerging upon approaching $T_c$. $F^{(i,z)}$ is the force per length of the hypercylinder due to its extension in the translationally invariant direction. In the spirit of a systematic expansion in terms of $\epsilon=4-d$ around the upper critical dimension we study the scaling functions $K$ within MFT as given by Eq.~\eqref{LGW-Hamiltonian} for hypercylinders in $d=4$, which captures the correct scaling functions in $d=4$ up to logarithmic corrections occurring in $d=d_c$~\cite{phasetransition_diehl,0295-5075-74-1-022}, which we do not take into account here. Since MFT renders the leading contribution to an expansion around $d=4$, geometrical configurations with small $D/R$, $L/R$, or $D/\xi_+$ are not expected to be described reliably by the present approach due to the dimensional crossover in narrow slit-like regions, which is not captured by the $\epsilon=4-d$ expansion.

The colloidal particles will also experience a lateral CCF
$F^{(i,x)}_{(a_1,a_2,b)}(D_1,D_2,R_1,R_2,L,t)$, for which it is convenient to
use the scaling form 

\begin{multline} \label{def_scaling_x}
F^{(i,x)}_{(a_1,a_2,b)}(D_1,D_2,R_1,R_2,L,t) = \\
= k_BT \frac{R_i}{L^{d-\mathfrak{D}+2}} K^{(i,x)}_{(a_1,a_2,b)}(\Xi,\Delta_1,\Delta_2,\Pi_1,\Pi_2)\,,
\end{multline}

\noindent
where $\Xi = {\rm sign}(t)L/\xi_\pm$ (i.e., $\Xi=L/\xi_+$ for $t>0$ and $\Xi=-L/\xi_-$ for $t<0$). Note that the choice of $\Xi$ as the scaling variable does not depend on the type of particle the force acts on. Equation \eqref{def_scaling_x} also describes the singular contribution to the lateral force near $T_c$. The total CCF acting on particle $i$ is

\begin{equation} \label{resulting_force}
\mathbf{F}^{(i,xz)}_{(a_1,a_2,b)} = F^{(i,x)}_{(a_1,a_2,b)}\mathbf{e}_x +
F^{(i,z)}_{(a_1,a_2,b)}\mathbf{e}_z ,
\end{equation}

\noindent
where $\mathbf{e}_x$ and $\mathbf{e}_z$ are the unit vectors pointing in $x$- and $z$-direction, respectively. Due to symmetry all other components of the CCF are zero.

As a reference configuration we consider a single spherical colloid of radius $R$ with BC $(a)$ at a surface-to-surface distance $D$ from a planar substrate with BC $(b)$. This colloid experiences (only) a normal CCF

\begin{multline} \label{def_scaling_norm}
F^{(*,z)}_{(a,b)}(D,R,T) =\\
= k_B T \frac{R}{D^{d-\mathfrak{D}+2}} K^{(*,z)}_{(a,b)}(\Theta={\rm sign}(t) \dfrac{D}{\xi_\pm},\Delta=\dfrac{D}{R})~.
\end{multline}

In the following we normalize the scaling functions $K^{(i,z)}_{(a_1,a_2,b)}$ and $K^{(i,x)}_{(a_1,a_2,b)}$ by the amplitude $K^{(*,z)}_{(+,+)}(\Theta=0,\Delta=1)$ of the CCF acting at $T_c$ on a single colloid for $(+,+)$ BCs at a surface-to-surface distance $D=R$. Accordingly, in the following we consider the normalized scaling functions

\begin{multline}  \label{def_normalized_scaling}
\overline{K}^{(i,s)}_{(a_1,a_2,b)}(\Lambda_s,\Delta_1,\Delta_2,\Pi_1,\Pi_2) =\\
= \frac{K^{(i,s)}_{(a_1,a_2,b)}(\Lambda_s,\Delta_1,\Delta_2,\Pi_1,\Pi_2)}{K^{(*,z)}_{(+,+)}(\Theta=0,\Delta=1)},
\end{multline}

\noindent
with $\Lambda_{s=z}=\Theta_i$ and $\Lambda_{s=x}=\Xi$. Experimentally it can be rather difficult to obtain $K^{(*,z)}_{(+,+)}(\Theta=0,\Delta=1)$. A standard alternative way to normalize is to take the more easily accessible amplitude $\Delta_{(+,+)}$ for the CCF at $T_c$ between two parallel plates with $(+,+)$ BCs, which is given within MFT by (see Ref.~\cite{trondle:074702} and references therein)

\begin{equation}
\Delta_{(+,+)} = -24\dfrac{[\mathcal{K}(1/\sqrt{2})]^4}{u} \simeq -283.61 u^{-1} ~,
\end{equation}

\noindent
where $\mathcal{K}$ is the elliptic integral of the first kind. Within MFT the amplitude $K^{(*,z)}_{(+,+)}(\Theta=0,\Delta=1)$ can be expressed in terms of $\Delta_{(+,+)}$:

\begin{equation}\label{estaequacaoetaoimportantequeelaganhouseupropriolabel}
K^{(*,z)}_{(+,+)}(\Theta=0,\Delta=1) \approx 0.4114 \Delta_{(+,+)} ~.
\end{equation}

\noindent
Equation \eqref{estaequacaoetaoimportantequeelaganhouseupropriolabel} allows for a practical implementation of the aforementioned normalization, which eliminates the coupling constant $u$, which is unknown within MFT.

We calculate the normal and lateral forces directly from the numerically determined order parameter profiles $\phi(x,y,z)$ by using the stress tensor which, within the Ginzburg-Landau approach, is given by~\cite{PhysRevE.56.1642,kondrat:204902,trondle:074702}

\begin{equation}
\mathcal{T}_{kl} = \dfrac{\partial\phi}{\partial r_k}\cdot \dfrac{\partial\phi}{\partial r_l} - \delta_{k,l} \left[ \dfrac{1}{2}(\nabla\phi)^2 + \dfrac{\tau}{2}\phi^2 + \dfrac{u}{4!}\phi^4 \right] ~,
\end{equation}

\noindent
with $\left\{ k,l\right\} \in \left\{ x,y,z\right\}$.

The first index of the stress tensor denotes the direction of a force, the second index denotes the direction of the normal vector of the surface upon which the force acts. Therefore one has

\begin{equation}
\dfrac{F^{(i,k)}_{(a_1,a_2,b)}}{k_{{\rm B}}T} = \dfrac{1}{\mathfrak{L}_{d-\mathfrak{D}}}\int_{A_i} \hat{n}_l \, \mathcal{T}_{kl} \, d^{\mathfrak{D}-1}r~,
\end{equation}

\noindent
where $A_i$ is a hypersurface enclosing particle $i$, $\hat{n}_l$ is the $l$-th component (to be summed over) of its unit outward normal, and $\mathfrak{L}_{d-\mathfrak{D}}=\int d^{d-\mathfrak{D}}r$ is the length of the $d-\mathfrak{D}$-dimensional hyperaxis of $\mathcal{C}_d$. In particular we focus on the normal and lateral CCFs acting on colloid (2) for the configuration shown in Fig.~\ref{system_sketch}, for $t\geqslant 0$, and with the binary liquid mixture at its critical concentration. In the following analysis we consider colloid (1) to be fixed in space at a sphere-surface-to-substrate distance $D_1=R_1$, equally sized colloids (i.e., $R_1=R_2$), and fixed $(b=+)$ BC for the substrate. We proceed by varying either the vertical ($z$-direction) or the horizontal ($x$-direction) position of colloid (2) by varying either $D_2$ or $L$, respectively. We also consider different sets of BCs for the colloids. In the following results the numerical error is typically less than $5\%$, unless explicitly stated otherwise.


\section{Results}\label{section_results}



\subsection{Total force}\label{subsection_total_force}

\begin{figure}
\begin{center}
\includegraphics[scale=0.5]{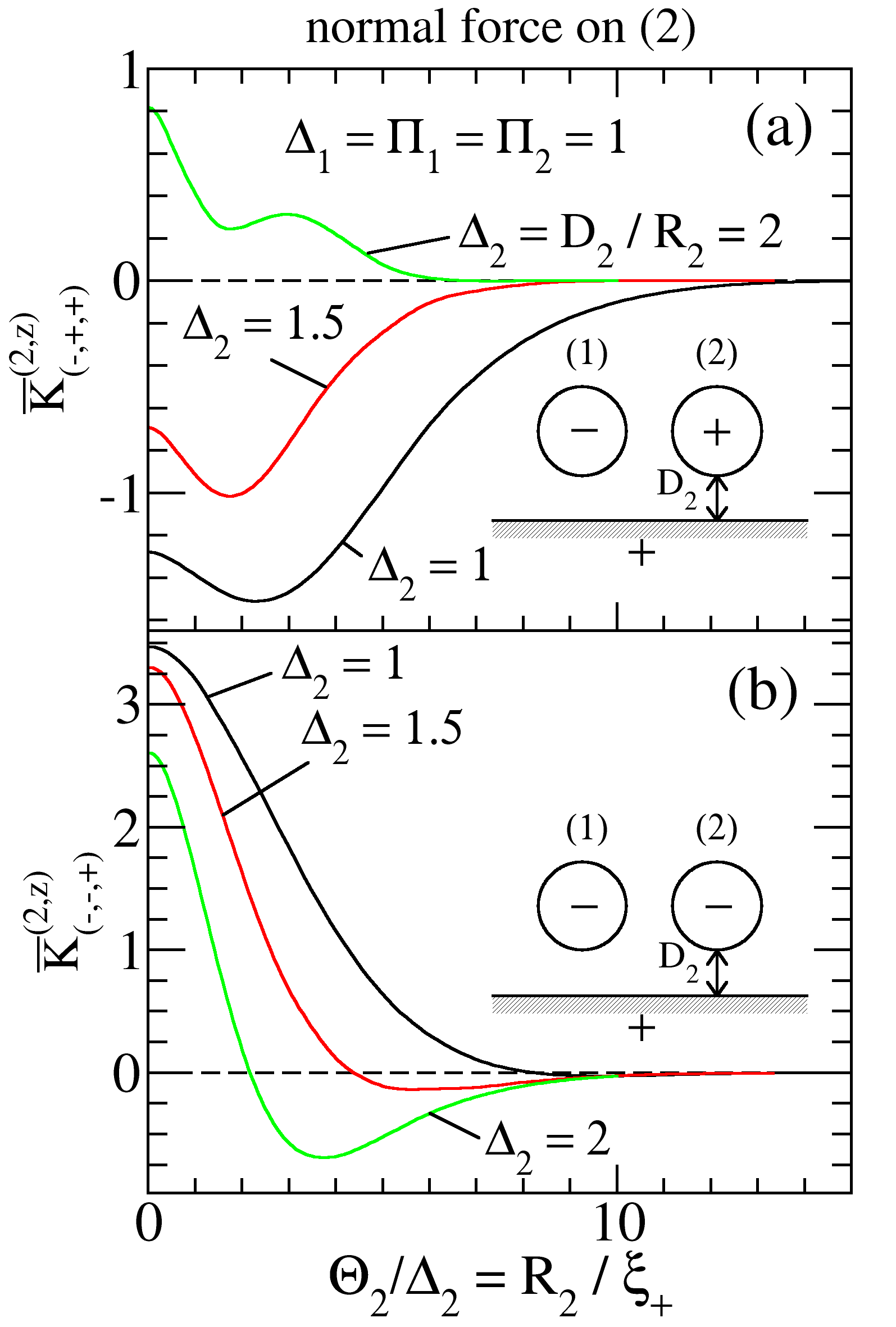}
\caption{
Normalized scaling functions
$\overline{K}^{(2,z)}_{(-,a_2,+)}(\Theta_2,\Delta_1=1,\Delta_2,\Pi_1=1, \Pi_2=1)$ of
the normal CCF acting on colloid (2) with BCs $(a_2=+)$ in (a) and $(a_2=-)$ in
(b). The scaling functions are shown for $t>0$ as functions of the scaling variable ratio $\Theta_2/\Delta_2 = R_2/\xi_+$ for three fixed values of the scaling variable $\Delta_2=D_2/R_2$: $\Delta_2=1$ (black lines), 1.5 (red lines), and 2 (green lines), while $\Delta_1=\Pi_1=\Pi_2=1$ for all curves in (a) and (b) so that $D_1=R_1=R_2=L$. For $R_2$ fixed the three curves correspond to three different vertical positions of colloid (2) with colloid (1) fixed in space. As expected, the forces become overall weaker upon increasing $\Delta_2$. $\overline{K}^{(2,z)}_{(-,a_2,+)}<0$ $(>0)$ implies that the colloid is
attracted to (repelled from) the substrate along the $z$-direction.} 
\label{D2_Z}
\end{center}
\end{figure}

In Fig.~\ref{D2_Z} we show the behavior of the normalized [Eq.~\eqref{def_normalized_scaling}] scaling function $\overline{K}^{(2,z)}_{(-,a_2,+)}(\Theta_2,\Delta_1=1,\Delta_2,\Pi_1=1,\Pi_2=1)$ of the normal CCF acting on colloid (2) with $(a_2=\pm)$ BCs close to a homogeneous substrate with $(b=+)$ BC and in the presence of colloid (1) with $(a_1=-)$ BC. The scaling functions are shown as functions of the scaling variable ratio $\Theta_2/\Delta_2=R_2/\xi_+$, i.e., for $t>0$. The various lines correspond to distinct fixed values of $\Delta_2$ as the sphere-surface-to-substrate distance in units of the sphere radius. Thus Fig.~\ref{D2_Z} shows the temperature dependence of the normal CCF on colloid 2 for three different values of $D_2$ and for colloid (1) fixed in space.

From Fig.~\ref{D2_Z} (a) one can see that, for colloid (1) with $(a_1= -)$ BC and for any given value of $\Theta_2/\Delta_2$, the scaling function of the normal CCF acting on colloid (2) with $(a_2 = +)$ BC changes sign upon varying $\Delta_2$. Due to the change of sign of $\overline{K}^{(2,z)}_{(-,+,+)}$, for any value of $\Theta_2/\Delta_2$ there is a certain value $\Delta^{(0)}_2 > 1$ at which the normal CCF acting on colloid (2) vanishes. For sphere-surface-to-substrate distances sufficiently large such that $\Delta_2 > \Delta^{(0)}_2$, colloid (2) is pushed away from the substrate due to the dominating repulsion between the two colloids in spite of the attraction by the substrate, whereas for $\Delta_2 < \Delta^{(0)}_2$ it is pulled to the substrate due to the dominating attraction between it and the substrate. This implies that, in the absence of additional forces, levitation of colloid (2) (i.e., zero total normal force) at height $D^{(0)}_2=\Delta^{(0)}_2 R_2$ is not stable against perturbations of the sphere-surface-to-substrate distance. On the other hand, upon varying temperature, any distance $D_2$ can become a stable levitation position for colloid (2) with $(a_2=-)$ BC in the presence of colloid (1) with $(a_1=-)$ BC [see Fig.~\ref{D2_Z} (b), according to which each scaling function corresponding to a certain value of $D_2$ exhibits a zero so that, at this zero, increasing (decreasing) $D_2$ at fixed temperature leads to an attraction (repulsion) to (from) the substrate]. In this case the attraction between the two colloids is dominating for large sphere-surface-to-substrate distances $\Delta_2$, while the repulsion between colloid (2) and the substrate dominates for small values of $\Delta_2$.


\begin{figure}
\begin{center}
\includegraphics[scale=0.52]{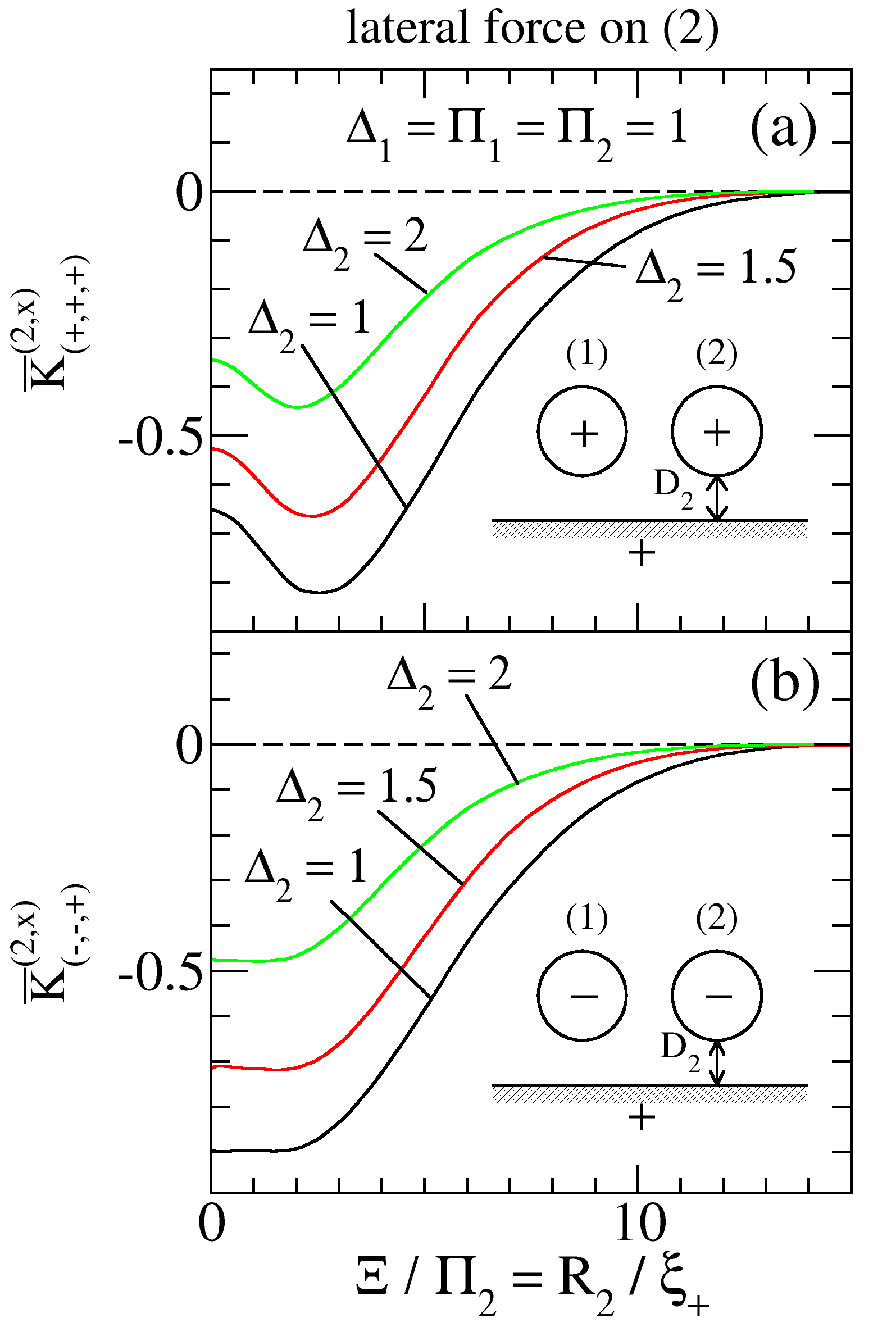}
\caption{
Normalized scaling functions
$\overline{K}^{(2,x)}_{(a_1,a_2,+)}(\Xi,\Delta_1=1,\Delta_2,\Pi_1=1,\Pi_2=1)$ of the lateral CCF acting on colloid (2) facing a homogeneous substrate with $(b=+)$ BC and in the presence of colloid (1) with $(a_1)=(a_2)$ BCs. The scaling functions are shown for $t>0$ as functions of the scaling variable ratio $\Xi/\Pi_2=R_2/\xi_+$ for three fixed values of the scaling variable $\Delta_2=D_2/R_2$: $\Delta_2=1$ (black curves), 1.5 (red curves), and 2 (green curves), while  $\Delta_1=\Pi_1=\Pi_2=1$ for all curves in (a) and (b) so that $D_1=R_1=R_2=L$. For $R_2$ fixed the three curves correspond to three different vertical positions of colloid (2) with colloid (1) fixed in space. As expected, the forces become overall weaker upon increasing $\Delta_2$. $\overline{K}^{\,(2,x)}_{(a_1,a_1,+)}<0$ implies that colloid (2) is attracted towards colloid (1). Two different sets of $(a_1,a_2,+)$ BCs are considered: $(+,+,+)$ in (a) and $(-,-,+)$ in (b).}
\label{D2_X}
\end{center} 
\end{figure}

Figure~\ref{D2_X} shows the behavior of the normalized scaling function
$\overline{K}^{(2,x)}_{(a_1,a_2,+)}(\Xi,\Delta_1=1,\Delta_2,\Pi_1=1,\Pi_2=1)$ of the lateral CCF acting on colloid (2) in the presence of colloid (1) having the same BC, i.e., $(a_1)=(a_2)=(\pm)$. The scaling functions are shown as functions of the scaling variable ratio $\Xi/\Pi_2=R_2/\xi_+$. From Figs.~\ref{D2_X} (a) and (b) one can infer that $\overline{K}^{\,(2,x)}_{(\pm,\pm,+)}<0$. Therefore colloid (2) is always attracted towards colloid (1) which has the same BC. Hence the substrate does not change the sign of the lateral CCF as compared with the attractive lateral CCF in the absence of the confining substrate. However, the shapes of the scaling functions for $(+,+,+)$ BCs [Fig.~\ref{D2_X} (a)] differ from the ones for $(-,-,+)$ BCs [Fig.~\ref{D2_X} (b)]; without the substrate, they are identical. In the former case and in contrast to the latter one, the scaling functions exhibit minima above $T_c$, which is reminiscent of the shape of the corresponding scaling functions in the absence of the substrate.


\begin{figure}
\begin{center}
\includegraphics[scale=0.45]{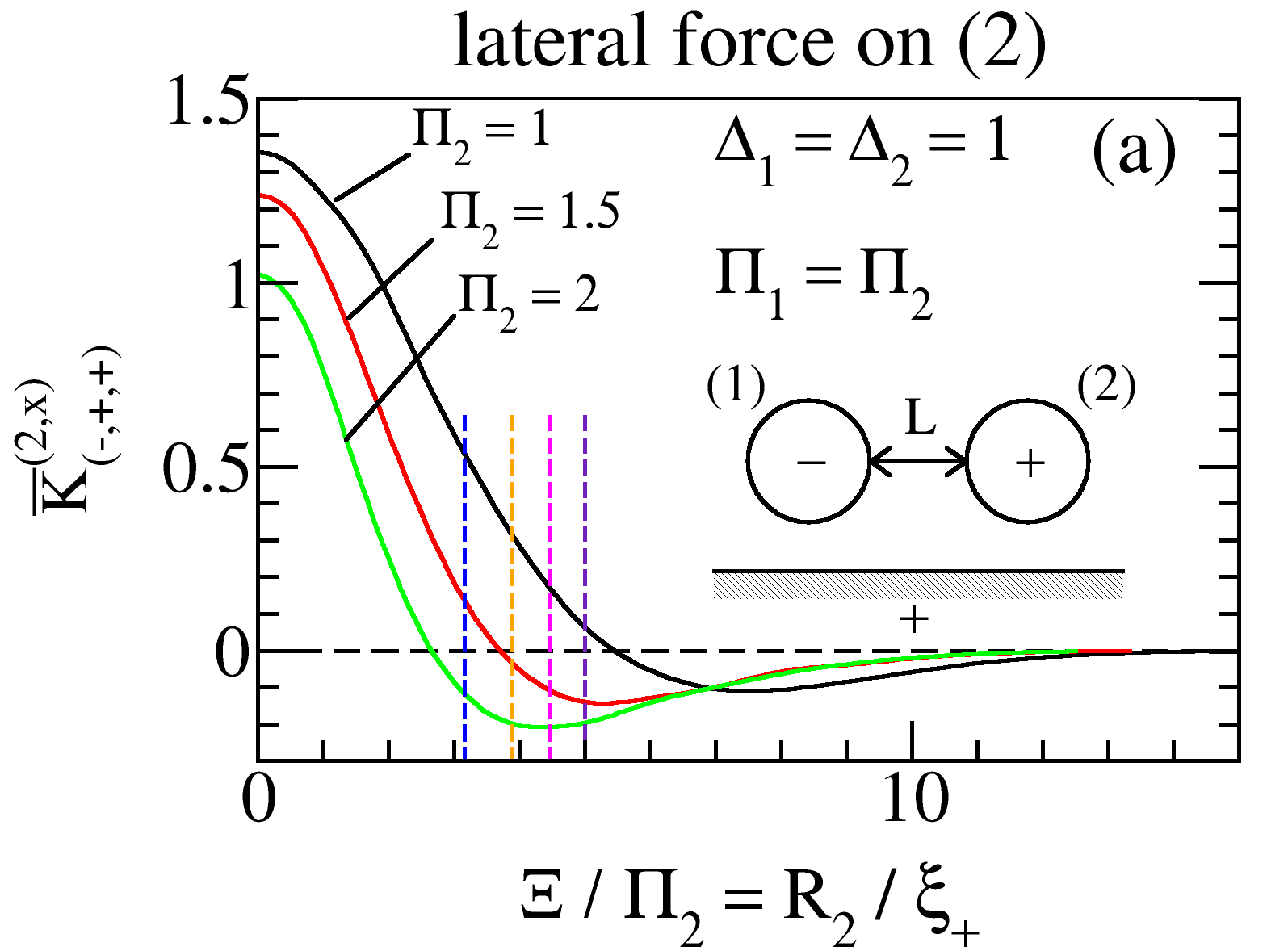}
\includegraphics[scale=0.45]{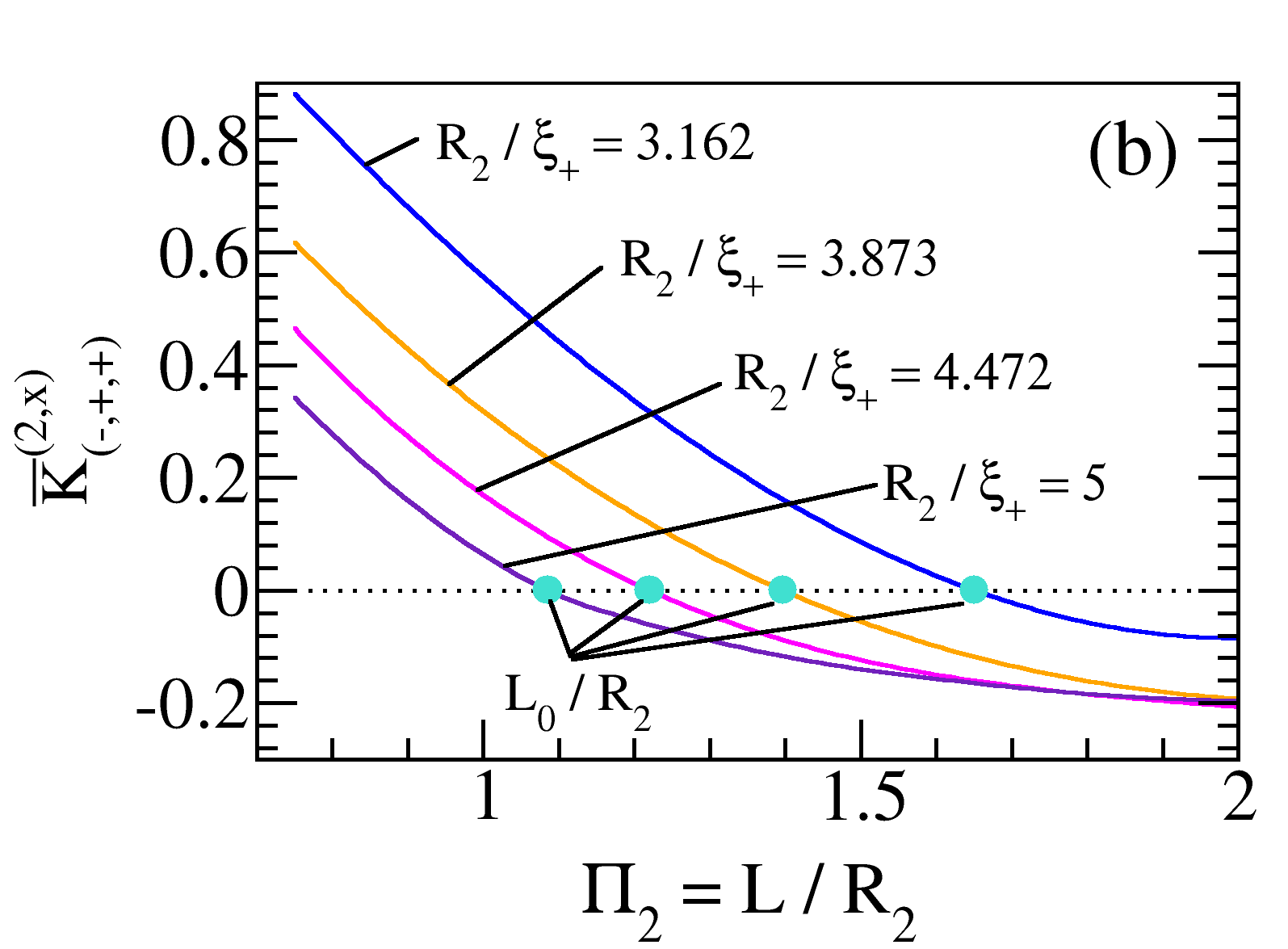} 
\includegraphics[scale=0.45]{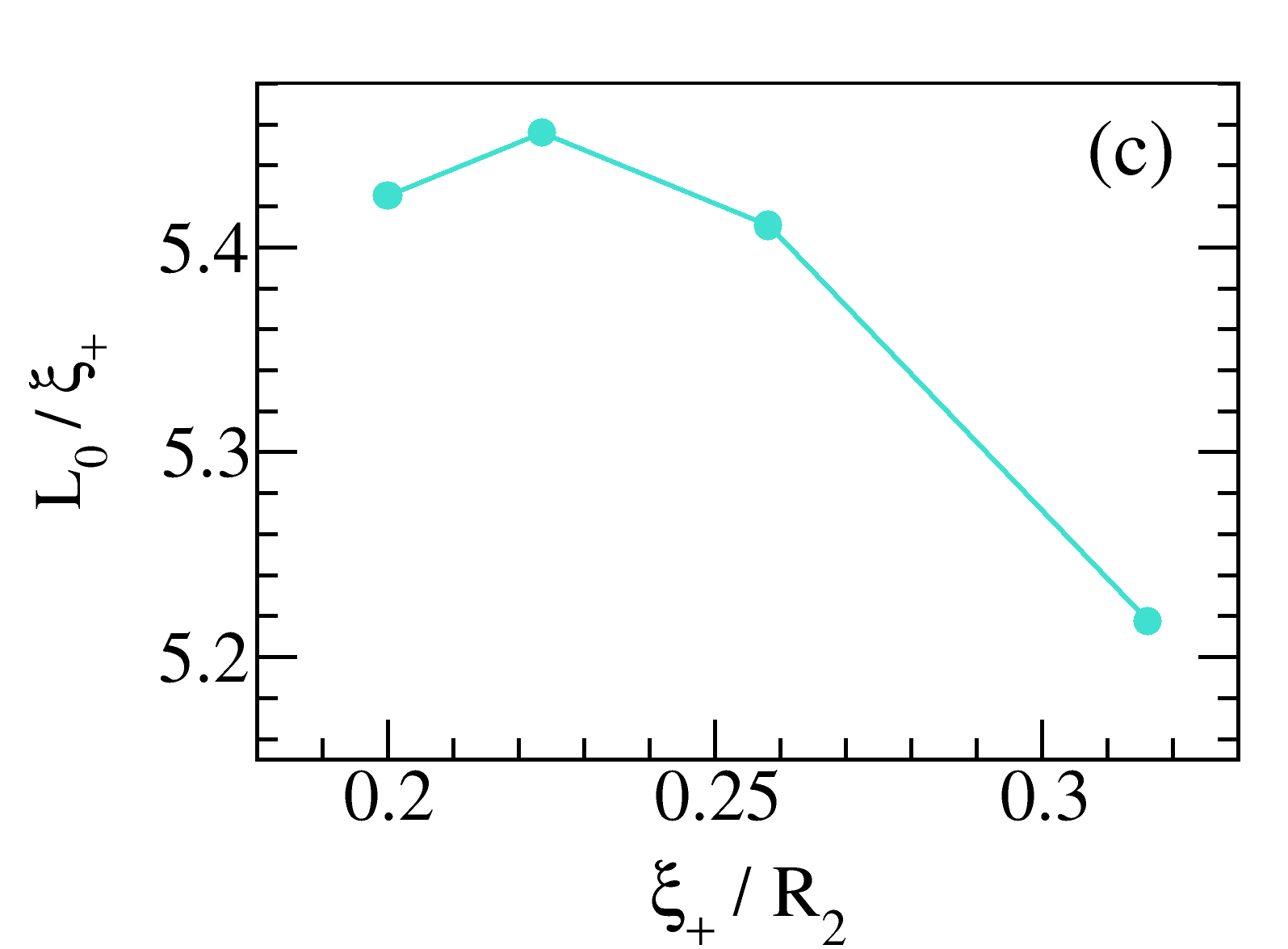}
\caption{
(a) Normalized scaling functions $\overline{K}^{(2,x)}_{(-,+,+)}(\Xi,\Delta_1=1,\Delta_2=1,\Pi_1=\Pi_2,\Pi_2)$ of the lateral CCF acting on colloid (2). Both colloids are taken to have the same size ($\Pi_1=\Pi_2=\Pi$) and the same sphere-surface-to-substrate distance ($\Delta_1=\Delta_2=1$). The scaling function is shown for $t>0$ as function of the scaling variable ratio $\Xi/\Pi_2=R_2/\xi_+$. $\overline{K}^{(2,x)}_{(-,+,+)}<0$ $(>0)$ implies that colloid (2) is attracted (repelled) by colloid (1). Black, red, and green curves correspond to lateral distances $\Pi_2=L/R_2=1$, 1.5, and 2, respectively, between the surfaces of the colloids [see Fig.~\ref{system_sketch}]. The zero $\Xi_0(\Pi_2)$ of $\overline{K}^{(2,x)}_{(-,+,+)}$ implies that for any lateral distance $L$ there is a reduced temperature such that for $\xi_+=\Xi_0(\Pi=L/R_2)L$ this distance represents an equilibrium lateral distance between the two colloids, provided they are located at equal sphere-surface-to-substrate distances. (b) $\overline{K}^{(2,x)}_{(-,+,+)}$ as function of the reduced surface-to-surface distance $\Pi_2=L/R_2$ between the two colloids for $\Xi/\Pi_2=R_2/\xi_+=3.162$ (blue line), 3.873 (yellow line), 4.472 (magenta line), and 5 (purple line). Each curve in (b) corresponds to the vertical dashed line with same color in panel (a). The change of sign of the scaling function (from positive to negative) as the distance between the colloids is increased indicates that the lateral CCF changes from repulsive to attractive, which means that there is a lateral position $L_0$ corresponding to a stable equilibrium point. In (c) we show how this equilibrium position, measured in units of $\xi_+$, varies as function of temperature (i.e., $\xi_+$) for fixed $R_2$. Note that $L_0$ is not proportional to $\xi_+$, since $L_0/\xi_+$ is not a constant. As a guide to the eye the four data points are connected by straight lines.}
\label{L_X}
\end{center} 
\end{figure}

In Fig.~\ref{L_X} we show the results obtained for the normalized scaling
functions $\overline{K}^{(2,x)}_{(-,+,+)}(\Xi,\Delta_1=1,\Delta_2=1,\Pi_1=\Pi_2, \Pi_2)$ of the lateral CCF acting on colloid (2). In Fig.~\ref{L_X}(a) the scaling function is shown as function of the scaling variable ratio $\Xi/\Pi_2=R_2/\xi_+$; the black, red, and green curves correspond to $L=R_2$, $L=1.5R_2$, and $L=2R_2$, respectively. In the absence of the substrate, the CCF between two colloids with opposite BCs is repulsive. However, as shown in Fig.~\ref{L_X}(a), in the presence of the substrate with $(b=+)$ BC, there is a change of sign in the scaling function of the lateral CCF. This implies that the lateral CCF acting between the two colloids changes from being attractive to being repulsive (or reverse) upon decreasing (increasing) the reduced temperature. Thus temperature allows one to control both the strength and the sign of the lateral CCF in the case of two colloids with opposite BCs being near a wall.

The at first sight unexpected lateral attraction between two colloids with opposite BCs in the presence of the substrate (i.e., $\overline{K}^{(2,x)}_{(-,+,+)}<0$) can be understood as follows. In the absence of the two colloids, the order parameter profile $m(\mathbf{r})$ is constant along any path within a plane $z=const$ because in this case $m(\mathbf{r}) = m(z)$. Since the substrate area is much larger than the surface areas of the colloids, one can regard the immersion of these colloidal spheres as a perturbation of this profile. In Figs.~\ref{L_X}(a) and (b), the region within which the scaling function is negative (corresponding to an attractive force) indicates that under these circumstances [i.e., when the colloids are sufficiently away from each other; see Fig.~\ref{L_X}(b)] the perturbation generated by the presence of the spheres decreases upon decreasing the lateral distance between them. This causes the colloids to move towards each other in order to weaken the perturbation by reducing its spatial extension; this amounts to an attraction, i.e., $\overline{K}^{(2,x)}_{(-,+,+)}<0$. On the other hand, when they are sufficiently close to each other the pairwise interaction between the two colloids dominates and the total lateral CCF is positive (i.e., repulsive). In Fig.~\ref{L_X}(c) we show how the equilibrium lateral distance $L_0$ measured in units of $\xi$ varies as function of temperature, i.e., $\xi$ for fixed $R$.


\begin{figure}
\begin{center}
\includegraphics[scale=0.7]{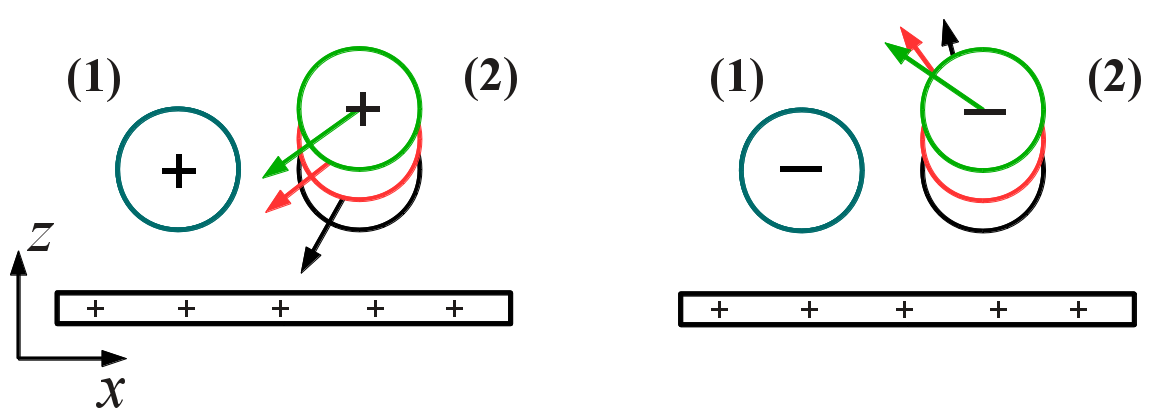}
\caption{
Sketch showing the unit vector pointing into the direction of the resulting CCF acting on colloid (2) for $(+,+,+)$ and $(-,-,+)$ BCs, with $\Delta_1 = \Pi_1 = \Pi_2 = 1$. The black rectangles represent the substrate and blue circles represent colloid (1) while black, red, and green circles represent colloid (2) with $\Delta_2 = 1, 1.5$, and 2, respectively. The centers of all colloids lie in the plane $y=0$.}
\label{direction_resulting_force_D2}
\end{center} 
\end{figure}

In order to determine the preferred arrangement of the colloids, we have also analyzed the direction of the total CCF $\mathbf{F}^{(2,xz)}_{(a_1,a_2,b)}$ acting on colloid (2) [see Eq.~\eqref{resulting_force}] for several spatial configurations and BCs. For $(+,+,+)$ BCs we have found that the colloids tend to aggregate laterally in such a way that several particles with $(+)$ BC, facing a substrate with the same BC, can be expected to form a monolayer on the substrate. On the other hand, for the case of $(-,-,+)$ BCs, we have found that the colloids can be expected to aggregate on top of each other so that a collection of colloids with such BCs is expected to form three-dimensional sessile clusters. These tendencies become more pronounced upon approaching the critical point (see Fig.~\ref{direction_resulting_force_D2}). Similar results have been found by Soyka et al.~\cite{PhysRevLett.101.208301} in experiments using chemically patterned substrates. For them the authors have found indeed that colloids with $(-)$ BC distributed over those parts of the substrate with the same BC [which is equivalent to $(+,+,+)$ BCs] aggregate and form a single layer. Moreover, they have found that colloids distributed over parts of the substrate with opposite BC [corresponding to $(-,-,+)$ BCs] form three-dimensional clusters.


\subsection{Many-body forces}\label{subsection_many-body_force}


We have determined the many-body force acting on particle $(i)$ by subtracting from
the total force $\mathbf{F}^{(i,xz)}_{(a_1,a_2,b)}$ [see Eq.~\eqref{resulting_force}] the sum of the pairwise forces acting on it, i.e., the colloid-colloid (CC) and the colloid-substrate (CS) forces. Accordingly the many-body CCF $\mathbf{F}^{(i,xz,MB)}_{(a_1,a_2,b)}(D_1,D_2,R_1,R_2,L,t)$ acting on colloid ($i$) is given by (see Fig.~\ref{system_sketch})

\begin{multline} \label{resulting_MB_force}
\mathbf{F}^{(i,xz,MB)}_{(a_1,a_2,b)} = \mathbf{F}^{(i,xz)}_{(a_1,a_2,b)}(D_1,D_2,R_1,R_2,L,t) -\\
\,\\
\shoveleft{-\mathbf{F}^{(i,xz,CC)}_{(a_1,a_2)}( D_1, D_2, R_1,R_2,L,t) -}\\
\,\\
\shoveleft{-\mathbf{F}^{(i,xz,CS)}_{(a_i,b)}(D_i,R_i,t)\,,}\\
\end{multline}


\noindent
where

\begin{widetext}
\begin{multline} \label{ss_force_definition}
\mathbf{F}^{(i,xz,CC)}_{(a_1,a_2,b)}(D_1, D_2, R_1,R_2,L,t) \equiv \mp \mathrm{sign}(a_1a_2)\left( \mathbf{e}_x \cos\alpha + \mathbf{e}_z \sin\alpha \right) \times \\
\times f_{(a_1,a_2)} \left( \sqrt{(D_2-D_1+R_2-R_1)^2 + (L+R_1+R_2)^2}-R_1-R_2, R_1, R_2, t \right)
\end{multline}
\end{widetext}

\noindent
with

\begin{equation}
\alpha = \arctan{\left( \dfrac{D_2-D_1+R_2-R_1}{L+R_1+R_2}\right) }
\end{equation}

\noindent
is the pairwise colloid-colloid force (acting on colloid 2 $(-)$ or 1 $(+)$ with 2 having the larger $x$-coordinate) expressed in terms of the absolute value $f_{(a_1, a_2)}(\mathfrak{u}, R_1, R_2, t )$ of the force between two colloids at surface-to-surface distance $\mathfrak{u}$ in free space. $\mathbf{F}^{(i,xz,CS)}_{(a_i,b)}(D_i,R_i,t)$ is the CCF between the substrate and a single colloid $(i)$.

We have studied both the normal

\begin{equation}
F^{(i,z,MB)}_{(a_1,a_2,b)} = \mathbf{F}^{(i,xz,MB)}_{(a_1,a_2,b)} \cdot \mathbf{e}_z
\end{equation}

\noindent
and the lateral

\begin{equation}
F^{(i,x,MB)}_{(a_1,a_2,b)} = \mathbf{F}^{(i,xz,MB)}_{(a_1,a_2,b)} \cdot \mathbf{e}_x
\end{equation}

\noindent
many-body CCFs which are characterized by corresponding scaling functions [compare Eqs. ~\eqref{def_scaling_z} and \eqref{def_scaling_x}]:

\begin{multline}\label{def_MB_scaling_z}
F^{(i,z,MB)}_{(a_1,a_2,b)} = \\
= k_BT \frac{R_i}{D_i^{d-\mathfrak{D}+2}} K^{(i,z,MB)}_{(a_1,a_2,b)}(\Theta_i,\Delta_1,\Delta_2,\Pi_1,\Pi_2),
\end{multline}

\noindent
and

\begin{multline}\label{def_MB_scaling_x}
F^{(i,x,MB)}_{(a_1,a_2,b)} = \\
= k_BT \frac{R_i}{L^{d-\mathfrak{D}+2}}K^{(i,x,MB)}_{(a_1,a_2,b)}(\Xi,\Delta_1,\Delta_2,\Pi_1,\Pi_2).
\end{multline}

\begin{figure}
\begin{center}
\includegraphics[scale=0.5]{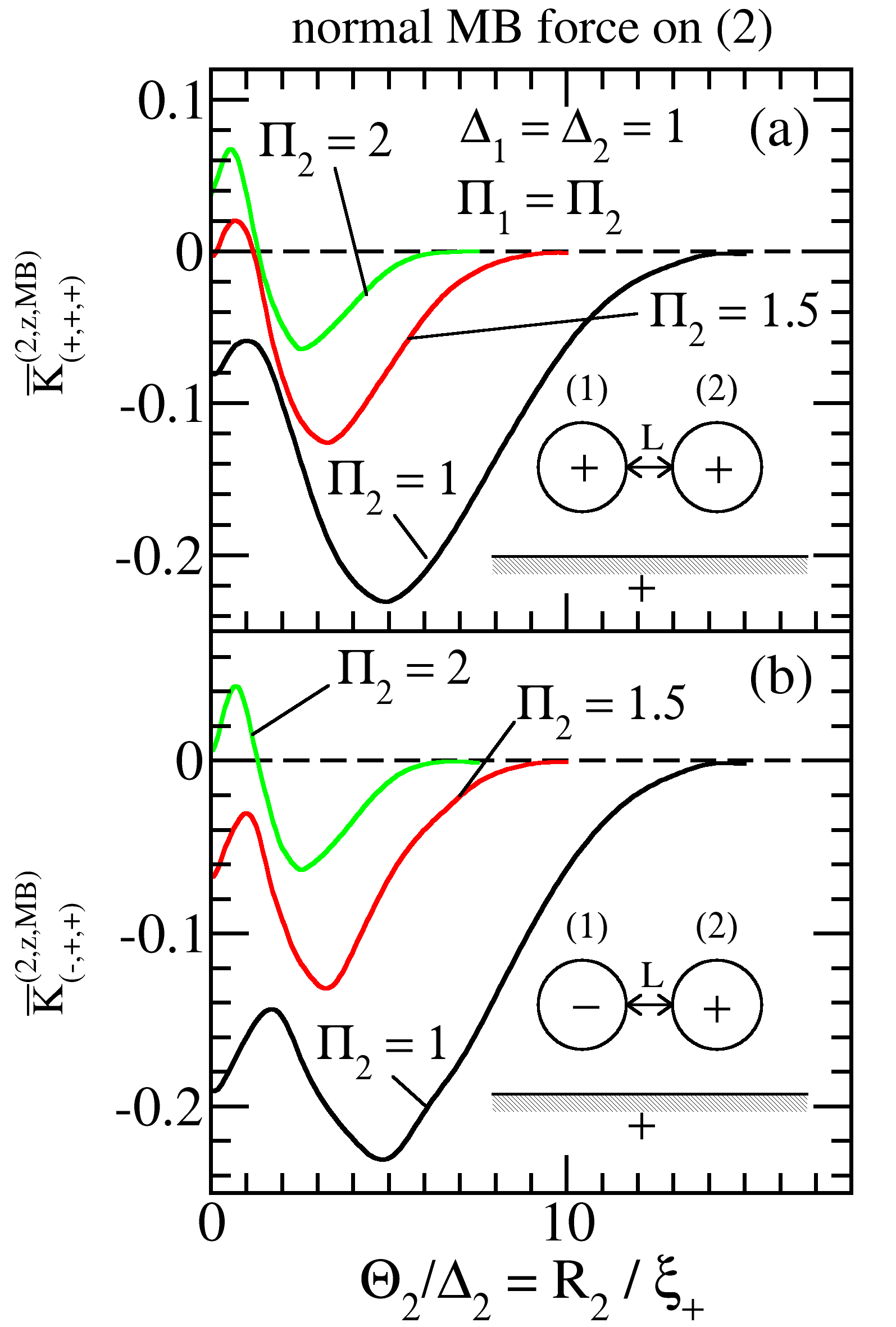}
\caption{
Normalized scaling functions
$\overline{K}^{(2,z,MB)}_{(a_1,+,+)}(\Theta_2,\Delta_1=1,\Delta_2=1,\Pi_1=\Pi_2, \Pi_2)$ of the many-body normal CCF acting on colloid (2) for $R_1=R_2=R$ and $D_1=D_2=R$. The scaling functions are shown as functions of the scaling variable ratio $\Theta_2/\Delta_2 = R/\xi_+$ for the sets of BCs $(+,+,+)$ in (a) and $(-,+,+)$ in (b). The black, red, and green lines correspond to $L/R=1, 1.5$, and $2$, respectively. Figures \ref{D2_Z}(a) and \ref{MB_L_Z}(b) allow a direct comparison between the full CCF and the corresponding many-body contribution (note the different scales of the ordinates.)}
\label{MB_L_Z}
\end{center} 
\end{figure}

In Fig.~\ref{MB_L_Z} we show the normalized [see Eqs.~\eqref{def_scaling_norm} and \eqref{def_normalized_scaling}] scaling functions $\overline{K}^{(2,z,MB)}_{(a_1,+,+)}(\Theta_2,\Delta_1=1,\Delta_2=1,\Pi_1=\Pi_2, \Pi_2)$ of the many-body normal CCF acting on colloid (2). This figure reveals similar results for $(+,+,+)$ [Fig.~\ref{MB_L_Z} (a)] and
$(-,+,+)$ [Fig.~\ref{MB_L_Z} (b)] BCs. In these cases, each MB scaling function exhibits both a maximum and at least one minimum, the former one appearing for smaller values of the scaling variable $R/\xi_+$ (i.e., at temperatures closer to $T_c$). For a certain range of temperatures close to $T_c$, as the distance $L$ between the colloids increases, the many-body normal CCF changes from attractive to repulsive. This shows that for each temperature within this range there is a lateral distance $L^{MB}_0/R$ for which the many-body contribution to the normal force acting on colloid (2) is zero. This means that under such conditions the sum of pairwise forces provides a quantitatively reliable description of the total force acting on colloid (2). For temperatures sufficiently far from $T_c$, the many-body normal CCF is always attractive with a monotonic dependence on $R/\xi_+$. Here, as in Figs.~\ref{D2_Z} - \ref{L_X}, the CCFs decay exponentially for $\xi_+\to 0$.

As expected, the many-body effects are more pronounced if the colloids are
closer to each other and/or closer to the substrate. Indeed for situations in which the colloids are close to each other [see, e.g., the black curves in Figs.~\ref{MB_L_Z} (a) and (b)] we have found that when the normal many-body CCF reaches its maximal strength, corresponding to the minimum of the scaling function $K^{(2,z,MB)}_{(a_1,+,+)}$ at $R/\xi_+\approx 5$, the relative contribution of the many-body CCF reaches $25\%$ of the strength of the total normal CCF. For larger distances between the colloids [see, e.g., the green curves in Figs.~\ref{MB_L_Z} (a) and (b)] this relative contribution is smaller (around $12\%$ for $\Pi_2=1.5$ and $6\%$ for $\Pi_2=2$).


We are not aware of results for the quantum-electrodynamic Casimir interactions which are obtained along the same lines as our CCF analysis above. Nonetheless, in order to assess the significance of our results we compare them with the results in Ref.~\cite{PhysRevA.83.042516}, which is the closest comparable study which we have found in the literature. Therein the authors study theoretically two dielectric spheres immersed in ethanol while facing a plate. Depending on the kind of fluid and on the materials of the spheres and of the plate as well as on the distances involved, also the quantum-electrodynamic Casimir force can be either attractive or repulsive. It is well known~\cite{PhysRevLett.89.033001,0305-4470-37-38-R01} that for two parallel plates with permittivities $\epsilon_a$ and $\epsilon_b$ separated by a fluid with permittivity $\epsilon_f$ and without further boundaries, the quantum-electrodynamic Casimir force is repulsive if $\epsilon_a<\epsilon_f<\epsilon_b$ within a suitable frequency range. In Ref.~\cite{PhysRevLett.104.160402} it is stated that this also holds for two spheres immersed in a fluid. The authors of Ref.~\cite{PhysRevA.83.042516} analyze the effect of nonadditivity for the above system by studying the influence of an additional, adjacent substrate on the equilibrium separation $d$ between two nanometer size dielectric spheres. To this end, they consider two spheres of different materials with the same radii $R_1=R_2=R$ and the same surface-to-plate distances $D_1=D_2=D$ and analyze how the lateral equilibrium distance $L_D$ between the spheres depends on $D$. By comparing the equilibrium distance $L_D$ with that in the absence of the substrate, $L_\infty$, they find that $L_D$ increases or decreases (depending on the kind of materials of the spheres) by as much as $15\%$ as the distance from the plate varies between $D=\infty$ and $D\approx R$. They also find that ``the sphere-plate interaction changes the sphere-sphere interaction with the same sign as $D$ becomes smaller'', which means that if the sphere-plate force is repulsive (attractive), the sphere-sphere one will become more repulsive (attractive) upon decreasing the distance from the plate $D$. By construction these changes are genuine many-body contributions. In the case of two chemically different spheres, the sign of the many-body force contribution (i.e., whether it is attractive or repulsive) agrees with the sign of the stronger one of the two individual sphere-plate interactions.

\begin{figure}
\begin{center}
\includegraphics[scale=0.65]{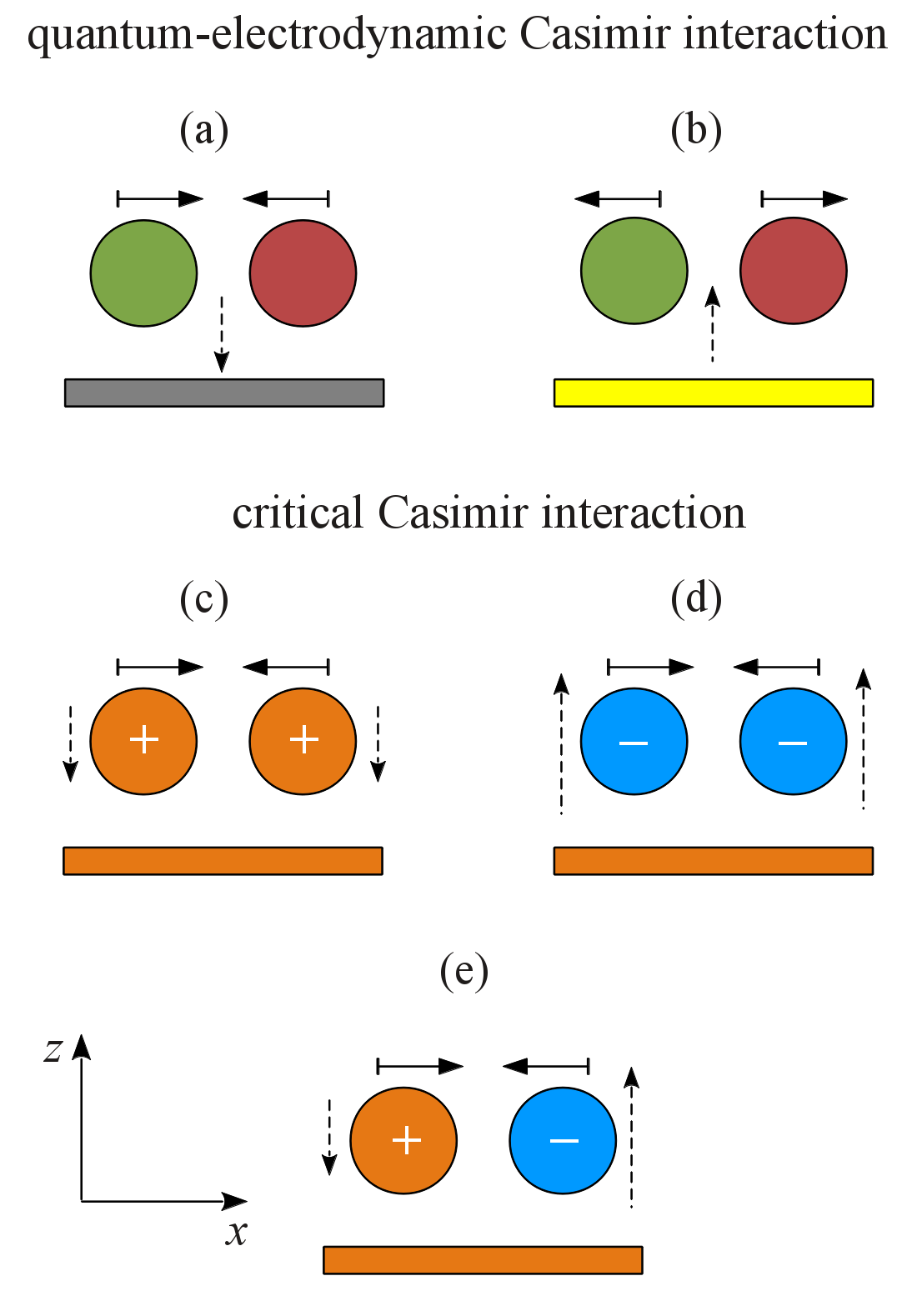}
\caption{
Comparison of the direction of the lateral many-body forces (solid arrows) due to quantum-electrodynamic and critical Casimir interactions. In the case of the quantum-electrodynamic interaction [(a) and (b)] and alluding to the system studied in Ref.~\cite{PhysRevA.83.042516}, the circles represent the projections of dielectric spheres with equal radii, the green one corresponding to polystyrene and the red one to silicon; the rectangles represent semi-infinite plates with their surfaces perpendicular to the $x-z$ plane, the gray and the yellow one being Teflon and gold, respectively. The system is immersed in ethanol, which is not indicated in the figure. In (a) and (b) each dashed arrow indicates the direction of the stronger one of the two corresponding pairwise forces between the dielectric spheres and the plate, which turns out to determine the direction of the many-body lateral force acting on the spheres: if the stronger one of the two pairwise forces is attractive [repulsive], the lateral many-body force will also be attractive [repulsive] (see Ref.~\cite{PhysRevA.83.042516}). Also in the case of the critical Casimir interaction [(c), (d), and (e)], the circles and rectangles represent projections of spherical colloids and of homogeneous substrates, respectively: orange and blue indicate $(+)$ and $(-)$ BCs, respectively. In (c) and (d) the pairwise normal forces between each of the two spheres and the substrate are equal: attractive in (c) and repulsive in (d). In (e) the two pairwise normal forces have opposite directions with the repulsive one being the stronger one~\cite{PhysRevE.80.061143,0295-5075-80-6-60009,PhysRevE.79.041142}. The corresponding dashed arrows have the same meaning as in (a) and (b). According to Fig.~\ref{MB_X} the many-body lateral CCFs are attractive for all three cases (c), (d), and (e). The comparison shows that the systems in (a) and (c) behave similarly. However, the behavior of system (b) has no counterpart for CCFs [see (d) and (e)]. In this figure all surface-to-surface distances equal the sphere radius, which in our notation corresponds to $\Pi_1=\Pi_2=\Delta_1=\Delta_2=1$.}
\label{comparing}
\end{center} 
\end{figure}

In Fig.~\ref{comparing} we show schematically the system considered in Ref.~\cite{PhysRevA.83.042516} [(a) and (b)] and the system considered here [(c), (d), and (e)]. For the quantum-electrodynamic Casimir effect, the dielectric spheres are represented by circles of equal radii, with the green one corresponding to a polystyrene sphere and the red one to a silicon sphere. The semi-infinite plates are represented by gray and yellow rectangles for Teflon and gold, respectively. The whole configuration is immersed in ethanol which, for simplicity, is not shown in the figure. The dashed arrows indicate the direction of the strongest of the two pairwise sphere-substrate forces, while the solid arrows indicate the direction of the lateral many-body force. The directions of the arrows in Figs.~\ref{comparing} (a) and (b) are chosen as to illustrate the findings in Ref.~\cite{PhysRevA.83.042516}, according to which the sign of the \textit{lateral} many-body force is the same as the one of the strongest \textit{normal} pairwise sphere-plate force: attractive in (a) and repulsive in (b).

\begin{figure}
\begin{center}
\includegraphics[scale=0.55]{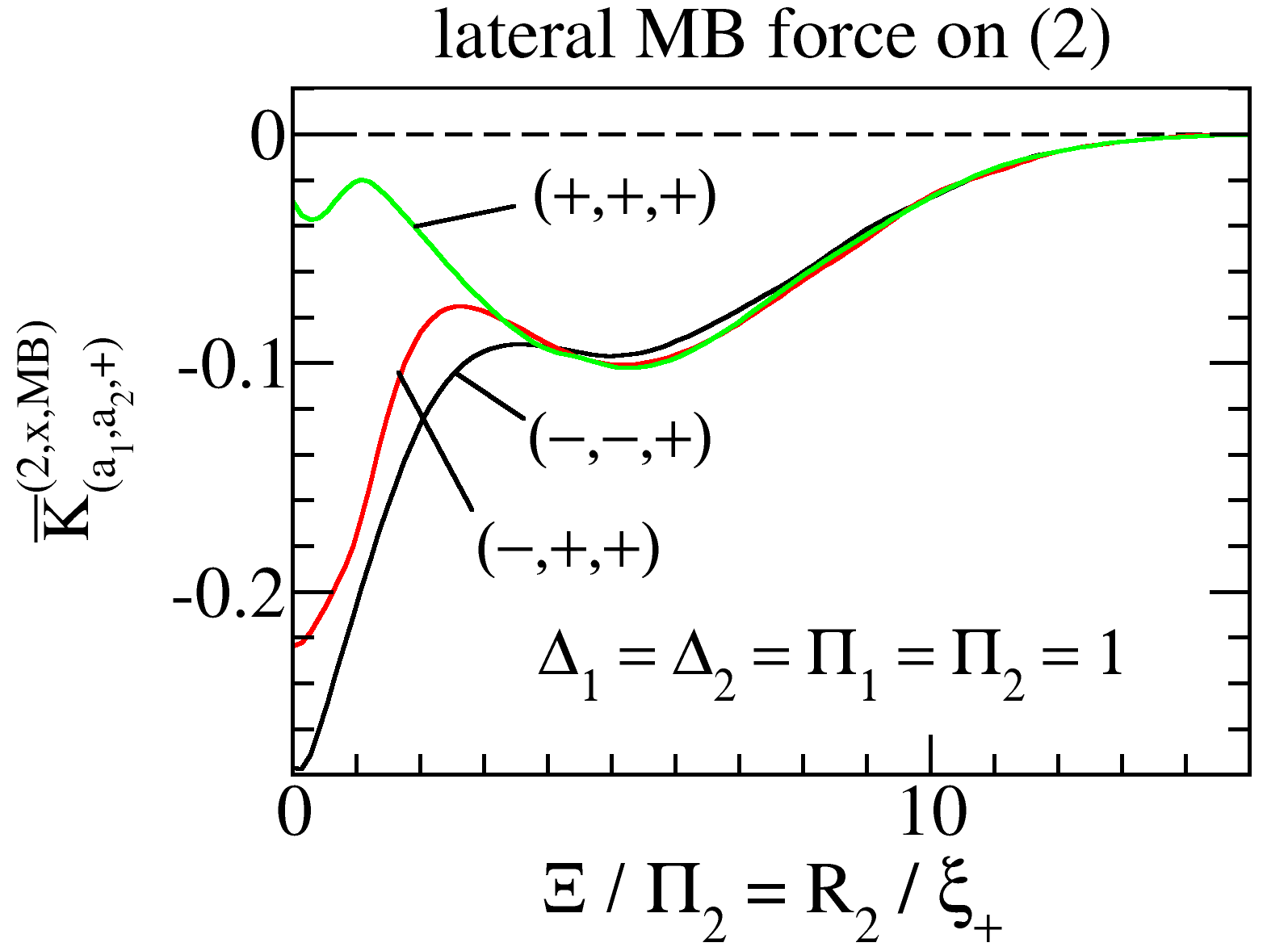}
\caption{
(a) Normalized scaling functions
$\overline{K}^{(2,x,MB)}_{(a_1,a_2,+)}(\Xi,\Delta_1=1,\Delta_2=1,\Pi_1=1, \Pi_2=1)$ of the lateral many-body CCF acting on colloid (2) for $R_1=R_2=R$ and $D_1=D_2=L=R$. The scaling function is shown as function of the scaling variable ratio $\Xi/\Pi_2 =\,R_2/\xi_+ = R/\xi_+$. The black, red, and green lines correspond to the BCs $(-,-,+)$, $(-,+,+)$, and $(+,+,+)$, respectively. For all three BCs the many-body contribution is not monotonic as function of temperature. Quantitatively the green, red, and black curves here should be compared with the black curves in Figs. \ref{D2_X}(a), \ref{D2_X}(b), and \ref{L_X}(a), respectively. However, for the data shown in \textit{this} figure the error bars (not shown) due to limits of the numerical accuracy are between 10$\%$ and 15$\%$. This is the main reason why we refrain from showing what would be an instructive plot such as $\overline{K}^{(2,x,MB)}_{(a_1,a_2,+)}$ as a function of $\Pi=\Pi_1=\Pi_2$ for various values of $\Delta=\Delta_1=\Delta_2$ and $\xi$ (as we did in Fig.~\ref{L_X}), which would allow for a direct comparison with the case of atoms.}
\label{MB_X}
\end{center} 
\end{figure}

Also in the case of the critical Casimir forces, depicted in Figs.~\ref{comparing} (c), (d), and (e), we represent the colloids by circles and the laterally homogeneous semi-infinite substrate by rectangles. The orange filling represents the $(+)$ BC while the blue filling represents the $(-)$ BC. Again, the dashed arrows indicate the direction of the stronger one of the two pairwise (normal) colloid-substrate forces, while the solid arrows indicate the direction of the lateral many-body contribution to the CCF. As one can infer from Fig.~\ref{MB_X}, the \textit{lateral} many-body CCF acting on colloid (2) is always attractive for the given geometrical configuration, regardless of the BCs.


We can also compare our results with those for two atoms close to the surface of a planar solid body. McLachlan~\cite{McLachlan1964} has tackled this problem by treating the solid as a uniform dielectric. By using the image method he derived an expression for the many-body corrections to the pairwise interaction energies, i.e., the atom-atom (London) and the atom-surface energies, in order to obtain the total interaction energy between the two atoms close to the surface. Qualitatively, he found that the leading contribution of the many-body correction leads to a repulsion if the atoms are side by side, i.e., at equal surface-to-substrate distances. 

Rauber et al.~\cite{PhysRevB.27.1314} used McLachlan's approach to study the electrodynamic screening of the van der Waals  interaction between adsorbed atoms and molecules and a substrate. The latter plays a role which is ``analogous to that of the third body in the three-body interaction between two particles embedded in a three-dimensional medium''. The van der Waals interaction between the two atoms at equal distances from the substrate is altered by the presence of the solid substrate and this perturbation is given by~\cite{McLachlan1964,PhysRevB.27.1314}

\begin{equation}\label{outraequacaosuperimportantegente}
\Delta V(\rho) = \frac{4C_{S1}}{\rho^6p^{3/2}}\left( \frac{1}{3} - \frac{l^2}{p\rho^2} \right) - \frac{C_{S2}}{\rho^6p^3}~,
\end{equation}

\noindent
where $\rho$ is the distance between the atoms, $l$ is the height above the image plane, which is the same for both atoms, and $p= 1 + 4l^2/\rho^2$. The coefficients are given by

\begin{equation}
C_{S1} = \frac{3\hslash}{\pi}\int_0^\infty \alpha^2(i\zeta)g(i\zeta)d\zeta~,
\end{equation}

\noindent
and

\begin{equation}
C_{S2} = \frac{3\hslash}{\pi}\int_0^\infty \alpha^2(i\zeta)g^2(i\zeta)d\zeta~,
\end{equation}

\noindent
with

\begin{equation}
g(i\zeta) = \left[ \epsilon(i\zeta) - 1 \right] / \left[ \epsilon(i\zeta) + 1 \right] ~,
\end{equation}

\noindent
where $\omega = i\zeta$ is an imaginary frequency, $\alpha (\omega)$ is the polarizability of the atoms (with the dimension of a volume), and $\epsilon (\omega)$ is the dielectric function of the solid (i.e., the substrate). The lateral force due to the perturbation potential given by Eq.~\eqref{outraequacaosuperimportantegente}, which is the analogue of the many-body contribution to the lateral CCF, follows from differentiating $\Delta V$ with respect to $\rho$:

\begin{multline}\label{nossaquelegaleumaequacaomaisimportantequeaoutra}
-\frac{d \Delta V}{d\rho} = -\frac{2}{\rho^7p^{3/2}}\left[ \frac{12l^2}{p\rho^2}\left( 2C_{S1} - \frac{C_{S2}}{p^{3/2}} \right) -\right. \\ \\
-\left. \frac{40l^4C_{S1}}{\rho^4p^2} -4C_{S1} +\frac{3C_{S2}}{p^{3/2}} \right] ~.
\end{multline}

\begin{figure}
\begin{center}
\includegraphics[scale=0.28]{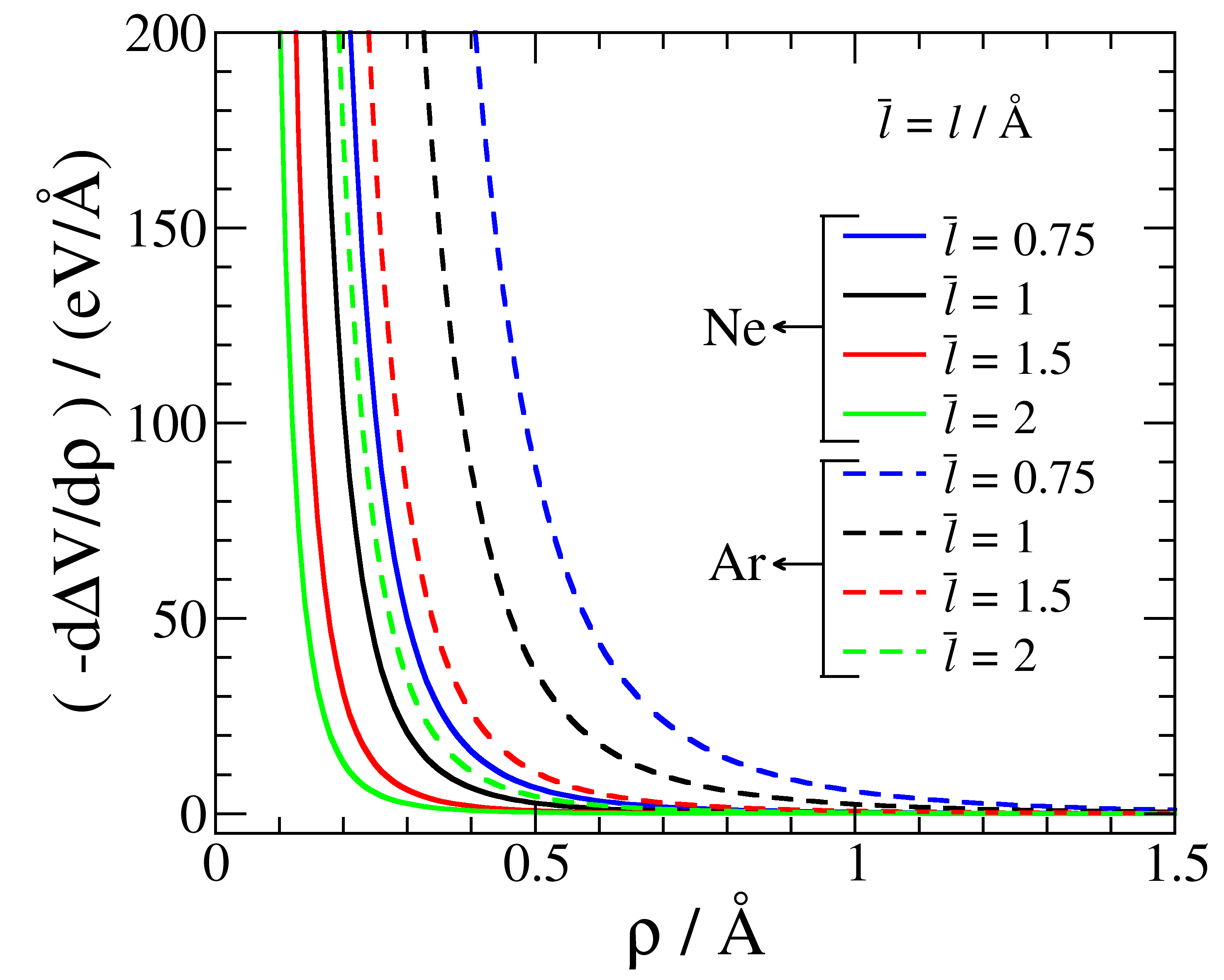}
\caption{
Many-body contribution to the lateral van der Waals force between two atoms adsorbed at a dielectric, homogeneous substrate [Eq.~\eqref{nossaquelegaleumaequacaomaisimportantequeaoutra}], derived from the expression for the excess potential given in Refs.~\cite{McLachlan1964} and ~\cite{PhysRevB.27.1314} [Eq.~\eqref{outraequacaosuperimportantegente}]. The forces are plotted as functions of the lateral separation $\rho$ between the two atoms for several equal vertical distances $l$ of the atoms from the substrate. The curves correspond to two sets of values for the coefficients $C_{S1}$ and $C_{S2}$ in Eq.~\eqref{outraequacaosuperimportantegente}~\cite{PhysRevB.27.1314}, corresponding to Ne (solid lines) and Ar (dashed lines).}
\label{mclachan_fig}
\end{center} 
\end{figure}

In Figure~\ref{mclachan_fig} we plot the lateral force given by Eq.~\eqref{nossaquelegaleumaequacaomaisimportantequeaoutra} as function of the distance $\rho$ between the two atoms for several (equal) distances $l$ above the substrate. We use the values provided in Ref.~\cite{PhysRevB.27.1314} for the coefficients $C_{S1}$ and $C_{S2}$: $C_{S1} = 1.33$ eV$\times$(\AA{})$^6$ and $C_{S2} = 0.70$ eV$\times$(\AA{})$^6$, which correspond to Ne, and $C_{S1} = 17.70$ eV$\times$(\AA{})$^6$ and $C_{S2} = 10.49$ eV$\times$(\AA{})$^6$, which correspond to Ar. As one can infer from Fig.~\ref{mclachan_fig}, the many-body contribution to the lateral van der Waals force is always repulsive and, as the two atoms approach the substrate, its strength increases. On the other hand, in the case of the many-body contribution to the lateral CCF, we have found that it is attractive for all BCs considered, if the surface-to-surface distances between the spheres and the sphere-surface-to-substrate distances are equal to each other and to the radius of the spheres (see Fig.~\ref{MB_X}).

Further, we can \textit{quantitatively} compare our results with those from Refs.~\cite{McLachlan1964} and ~\cite{PhysRevB.27.1314}. To this end, we assign values to the geometrical parameters characterizing the configuration of the two atoms close to the substrate and compare the results of Refs.~\cite{McLachlan1964} and ~\cite{PhysRevB.27.1314} with those for similar configurations in our model. For example, estimating the many-body contribution to the lateral van der Waals force for a configuration of two atoms close to a substrate corresponding to the configuration associated with the black curve~\footnote{In Fig.~\ref{MB_X} the black curve corresponds to a function of temperature but there is no temperature dependence of the van der Waals force between the atoms. Therefore the comparison has to be carried out by choosing a certain value of $R_2/\xi_+$}. in Fig.~\ref{MB_X} (i.e., $L/R=D/R=1$ in the case of the CCF and $l/\rho=1$ in the case of the two atoms), one obtains from Eq.~\eqref{nossaquelegaleumaequacaomaisimportantequeaoutra} a value for the relative contribution of the many-body force to the lateral force which corresponds to ca. $15\%$. This is comparable with the relative contribution of the many-body force to the lateral CCF sufficiently close to $T_c$, although in the case of the van der Waals force it is repulsive (Fig.~\ref{mclachan_fig}) while in the case of the CCF (Fig.~\ref{MB_X}) it is attractive.

Considering the decay of the many-body contribution to the normal CCF as function of the surface-to-surface distance $L$ between the spheres (for $D_1=D_2=D$), we can compare the corresponding decay of the normal many-body force $-d \Delta V/dl$ given by the potential $\Delta V$ in Eq.~\eqref{outraequacaosuperimportantegente}. For small separations $\rho$ between the atoms the many-body contribution to the normal van der Waals force increases as $\rho^{-3}$, while for large separations it decays as $\rho^{-8}$. By analyzing the data shown in Fig.~\ref{MB_L_Z} one finds that for $3<R_2/\xi_+<10$ the scaling function of the many-body contribution to the normal CCF decays slower than $L^{-2}$. This means that in this temperature regime the many-body contribution to the normal CCF is much more long ranged than the corresponding contribution to the normal van der Waals force in the case of two atoms close to a surface. For fixed $\rho$ the many-body contribution to the lateral van der Waals force decays as $l^{-3}$ upon increasing the distance of both atoms from the substrate whereas the normal force on a single atom decays as $l^{-4}$.

As a final remark we point out that we have not found a completely stable configuration for the two colloids (Fig.~\ref{system_sketch}): whenever there is a stable position in the horizontal (vertical) direction, the force is nonzero in the vertical (horizontal) direction. For example, consider a vertical path with $R/\xi_+=5$ in Figs.~\ref{D2_Z}(b) and \ref{D2_X}(b). From the first one can see that along this path the normal CCF changes from being repulsive to being attractive as the sphere-surface-to-substrate distance for colloid (2) is increased, implying that there is a vertical position of colloid (2) in which the normal CCF is zero. However, according to Fig.~\ref{D2_X}(b) the lateral CCF is always attractive regardless of the vertical position of colloid (2). This means that there is a configuration which is stable only in the normal direction. Accordingly, a dumbbell configuration with a rigid thin fiber between the two colloids can levitate over the substrate. Whether this configuration is stable with respect to a vertical tilt remains as an open question.


\subsection{Vertical alignment}\label{section_vertical}

\begin{figure}
\begin{center}
\includegraphics[scale=0.7]{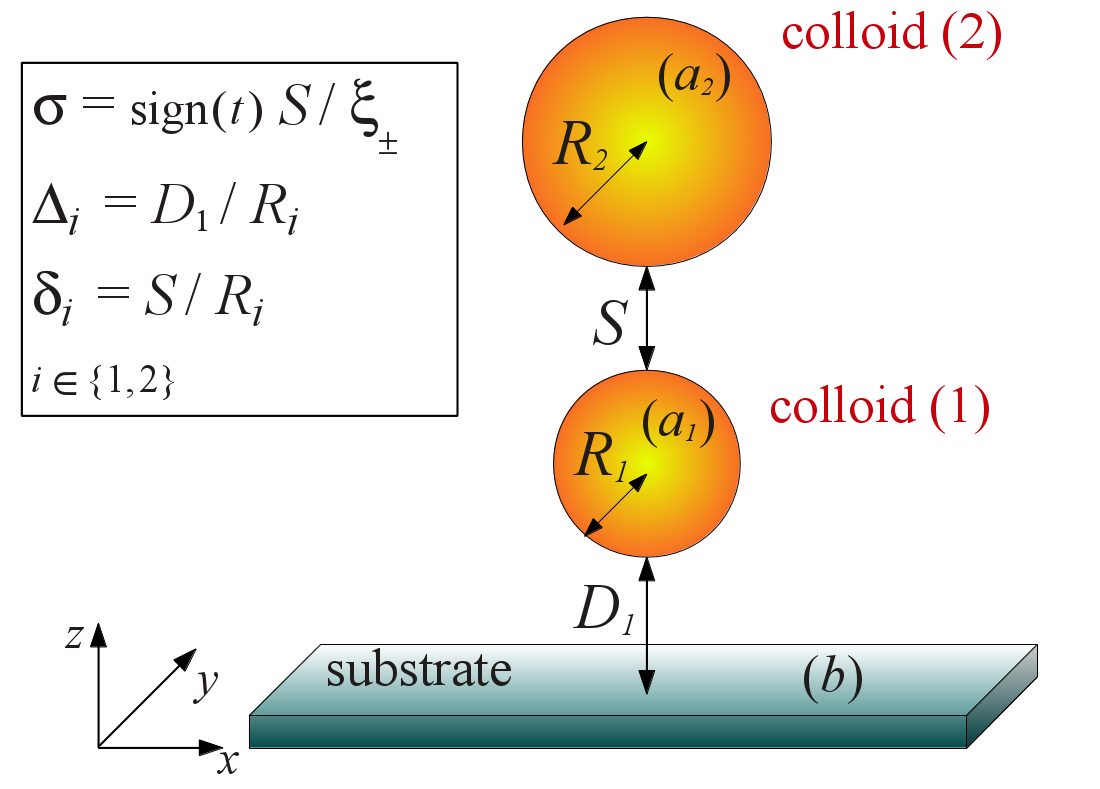}
\caption{
Two spherical colloidal particles of radii $R_1$ and $R_2$ immersed in a near-critical binary liquid mixture (not shown) and close to a homogeneous, planar substrate at $z=0$. The colloidal particle (1) with BC $(a_1)$ is located vertically at the sphere-surface-to-substrate distance $D_1$, whereas the colloidal particle (2) with BC $(a_2)$ is located vertically at the surface-to-surface distance $S$ between the spheres. The substrate exhibits BC $(b)$. The vertical distance between the centers of the spheres along the $z$-direction is given by $R_1 + S +R_2$, while the centers of both spheres lie on the vertical axis $x=0=y$. In the case of four spatial dimensions the figure shows a three-dimensional cut of the system, which is invariant along the fourth direction, i.e., the spheres correspond to parallel hypercylinders with one translationally invariant direction, which is $r_4$ [see Eq.~\eqref{eu<3hipercilindros}].}
\label{vertical_sketch}
\end{center} 
\end{figure}

We now turn our attention to the case in which the colloids are vertically aligned with respect to a planar, homogeneous substrate, i.e., when their centers have the same coordinates in both the $x$ and the $y$ direction (see Fig.~\ref{vertical_sketch}). We focus on the normal CCF acting on colloid (1) when the system is immersed in a near-critical binary liquid mixture at its critical concentration. As before we consider $\pm$ BCs corresponding to a strong adsorption preference for one of the two components of the confined liquid. In particular, we consider two three-dimensional spheres of radii $R_1$ and $R_2$ with BCs $(a_1)$ and ($a_2$), respectively, facing a laterally homogeneous substrate with BC $(b)$. Colloid (1) is positioned at a sphere-surface-to-substrate distance $D_1$ and colloid (2) is at a surface-to-surface distance $S$ from colloid (1) (see Fig.~\ref{vertical_sketch}). The coordinate system $(x,y,z)$ is chosen such that the centers of the spheres are located at $(0,0,D_1+R_1)$ and $(0,0,D_1+2R_1+S+R_2)$ so that the distance between the centers, along the $z$-axis, is given by $R_1 + S +R_2$. As before, the BCs of the system as a whole are represented by the set $(a_1,a_2,b)$, where $a_1$, $a_2$, and $b$ can be either $+$ or $-$.


The normal CCF $F^{(1,z)}_{(a_1,a_2,b)}(D_1,S,R_1,R_2,t)$ acting on colloid (1) along the $z$-direction takes the scaling form

\begin{multline} \label{def_scaling_vertical}
F^{(1,z)}_{(a_1,a_2,b)}(D_1,S,R_1,R_2,t) = \\
= k_BT \frac{R_1}{S^{d-\mathfrak{D}+2}}K^{(1,z)}_{(a_1,a_2,b)}(\sigma,\Delta_1,\Delta_2,\delta_1)\,,
\end{multline}

\noindent
where $\sigma={\rm sign}(t) S/\xi_\pm$ (i.e., $\sigma = S/\xi_+$ for $t>0$ and $\sigma = -S/\xi_-$ for $t<0$), $\Delta_1=D_1/R_1$, $\Delta_2=D_1/R_2$, and $\delta_1=S/R_1$; $\delta_2=S/R_2=\delta_1\Delta_2/\Delta_1$. Equation \eqref{def_scaling_vertical} describes the singular contribution to the normal force emerging upon approaching $T_c$. $F^{(1,z)}$ is the force on a hypercylinder divided by its extension in the translationally invariant direction [see Eq.~\eqref{eu<3hipercilindros}]. We use the same reference system as the one described by Eq.~\eqref{def_scaling_norm} in order to normalize the scaling function defined in Eq.~\eqref{def_scaling_vertical} according to

\begin{multline}\label{def_normalized_scaling_vertical}
\overline{K}^{(1,z)}_{(a_1,a_2,b)}(\sigma,\Delta_1,\Delta_2,\delta_1) = \\
=\frac{K^{(1,z)}_{(a_1,a_2,b)}(\sigma,\Delta_1,\Delta_2,\delta_1)}{K^{(*,z)}_{(+,+)}(\Theta=0,\Delta=1)}\,.
\end{multline}

We calculate the many-body contribution to the normal CCF acting on particle (1) $F^{(1,z,MB)}_{(a_1,a_2,b)}(D_1,S,R_1,R_2,t)$ by subtracting from the total force the sum of the pairwise forces acting on it, i.e., the colloid-colloid and the colloid-substrate forces [see Eq.~\eqref{resulting_MB_force}]. This many-body force takes the scaling form

\begin{multline}\label{def_MB_vertical_scaling_z}
F^{(1,z,MB)}_{(a_1,a_2,b)} = \\
= k_BT \frac{R_1}{S^{d-\mathfrak{D}+2}}K^{(1,z,MB)}_{(a_1,a_2,b)}(\sigma,\Delta_1,\Delta_2,\delta_1)\,.
\end{multline}

\begin{figure*}
\begin{center}
\includegraphics[scale=0.4]{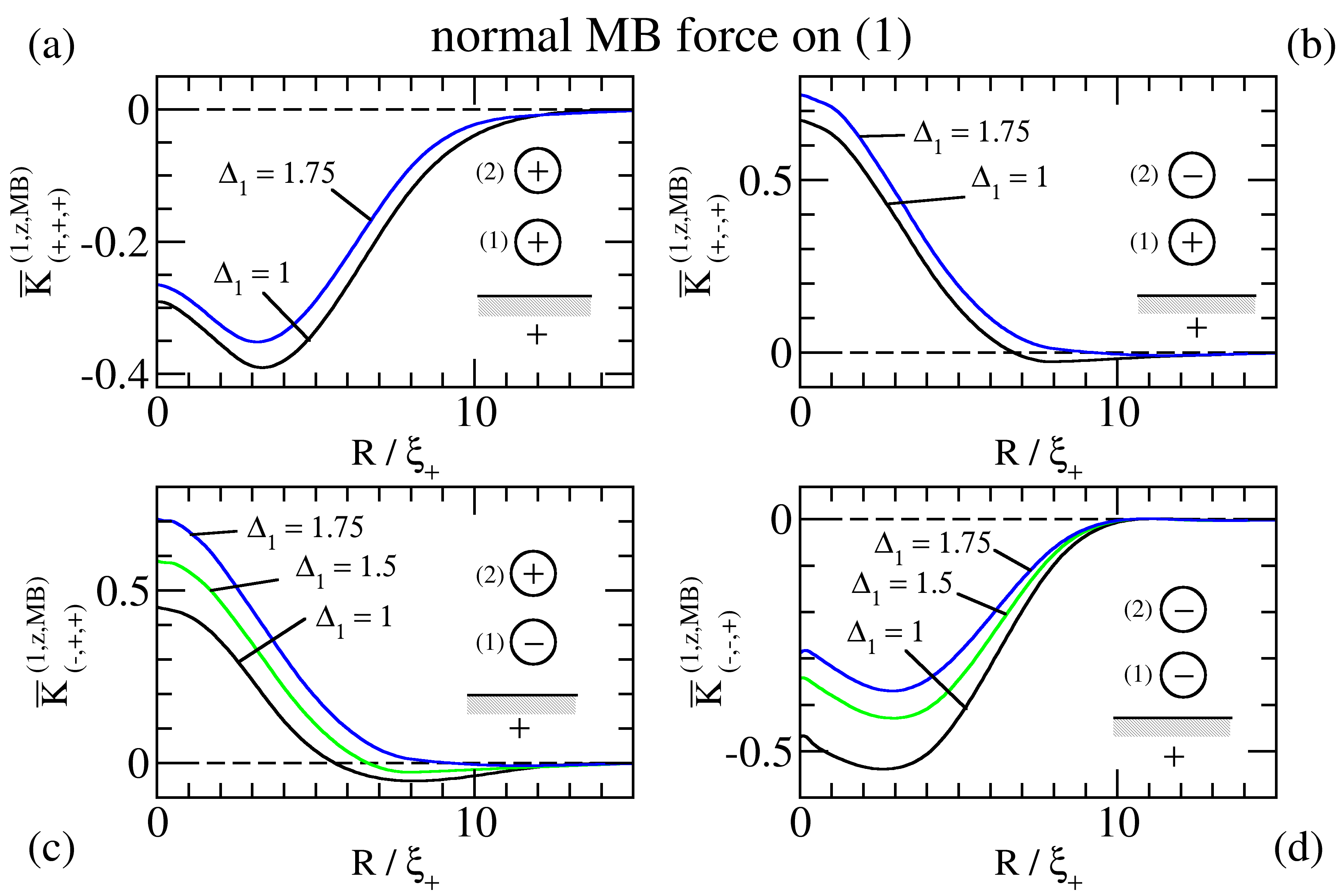}
\caption{
Normalized scaling functions
$\overline{K}^{(1,z,MB)}_{(a_1,a_2,+)}(\sigma,\Delta_1, \Delta_2=\Delta_1,\delta_1=1)$ of the many-body contribution to the normal CCF acting on colloid (1) for $R_1=R_2=R$ and $S=R$ (see Fig.~\ref{vertical_sketch}). The scaling functions are shown as functions of the scaling variable ratio $\sigma/\delta_1 = R_1/\xi_+$ for the sets of BCs $(+,+,+)$ in (a), $(+,-,+)$ in (b), $(-,+,+)$ in (c), and $(-,-,+)$ in (d). The black, green, and blue lines correspond to $D_1/R=1, 1.5$, and $1.75$, respectively. For $R_1$ fixed, different curves correspond to distinct vertical positions of the colloids (1) and (2) with unchanged surface-to-surface distance $S$ between the spheres. The apparent change of sign in Figs.~\ref{MB_vertical_Z}(b) and (c) is likely to be an artifact occurring within the error bars (which can be up to $10\%$ for the pure many-body CCF) due to numerical imprecision.}
\label{MB_vertical_Z}
\end{center}
\end{figure*}

In Fig.~\ref{MB_vertical_Z} we show the normalized [see Eqs.~\eqref{def_scaling_norm}, \eqref{def_normalized_scaling_vertical}, and \eqref{def_MB_vertical_scaling_z}] scaling functions $\overline{K}^{(1,z,MB)}_{(a_1,a_2,+)}(\sigma,\Delta_1, \Delta_2=\Delta_1,\delta_1=1)$ of the many-body contribution to the normal CCF acting on colloid (1) as functions of the scaling variable ratio $\sigma/\delta_1 = R_1/\xi_+$ for two spherical colloids of the same size ($R_1=R_2=R$). Keeping the surface-to-surface distance between the spheres fixed at $S=R$, we vary the sphere-surface-to-substrate distance $D_1$ for several BCs: $(+,+,+)$ in (a), $(+,-,+)$ in (b), $(-,+,+)$ in (c), and $(-,-,+)$ in (d).

From Figs.~\ref{MB_vertical_Z}(a) and (d) one can infer that if the colloids have symmetric BCs, the scaling function of the many-body normal CCF acting on colloid (1) is negative (i.e., it is directed towards the substrate) for any value of $R/\xi_+$. On the other hand, for non-symmetric BCs between the colloids [Figs.~\ref{MB_vertical_Z}(b) and (c)], the many-body contribution to the normal CCF acting on colloid (1) is positive for any value of $R/\xi_+$. The apparent change of sign in Figs.~\ref{MB_vertical_Z}(b) and (c) is likely to be an artifact occurring within the error bars due to numerical imprecision. The relative contribution of the many-body CCF to the total force is between 10$\%$ and 15$\%$.

We point out that this configuration with the the two colloids vertically aligned with respect to the substrate allows for a wide range of interesting aspects which will be further explored in future works.


\section{Conclusions and discussion}\label{section_conclusion}


We have investigated critical Casimir forces (CCFs) for a system composed of two equally sized spherical colloids ($R_1=R_2=R$) immersed in a near-critical binary liquid mixture and close to a laterally homogeneous substrate with $(b=+)$ boundary condition (BC) (see Fig.~\ref{system_sketch}). By denoting the set of BCs of the system as $(a_1,a_2,b)$, where $a_i$ corresponds to the BC at colloid $i=1,2$ and $b$ to the BC at the substrate, we have first focused on the total normal and lateral forces acting on one of the colloids [labeled as ``colloid (2)''] for several geometrical configurations of the system and various combinations of BCs at the colloids. Both the normal and the lateral forces are characterized by universal scaling functions [Eqs.~\eqref{def_scaling_z} and \eqref{def_scaling_x}, respectively], which have been studied in the one-phase region of the solvent as functions of $R_2/\xi_+$ and $L/R_2$. $L$ is the surface-to-surface distance between the two colloids, and $\xi_+$ is the bulk correlation length of the binary mixture in the mixed phase. We have used mean-field theory together with a finite element method in order to calculate the order parameter profiles, from which the stress tensor renders the normalized scaling functions associated with the CCFs.

For the scaling function of the total normal CCF acting on colloid (2) with $(a_2=+)$ BC, in the presence of colloid (1) with $(a_1 = -)$, we have found (Fig.~\ref{D2_Z}(a)) that the scaling function changes sign for a fixed value of $R_2/\xi_+$ as the distance $D_2$ between colloid (2) and the substrate increases, signaling the occurrence of an unstable mechanical equilibrium configuration of a vanishing normal force. For the total normal CCF acting on colloid (2) with $(a_2=-)$ BC and in the presence of colloid (1) with $(a_1=-)$, we have found (Fig.~\ref{D2_Z}(b)) that the force changes sign upon changing the temperature. For this combination of BCs, the equilibrium configuration of colloid (2) is stable in the normal direction.

Without a substrate, at the critical composition of the solvent the CCF between two $(+)$ spheres is identical to the one between two $(-)$ ones. This degeneracy is lifted by the presence of the substrate as one can infer from the comparison of the scaling functions for the lateral CCFs for $(+,+,+)$ and $(-,-,+)$ BCs [see Figs.~\ref{D2_X}(a) and (b), respectively]. In the first case, the shape of the scaling function resembles that of the two colloids far away from the substrate, with a minimum at $R_2/\xi_+\approx 2.5$. In the second case this minimum at $T>T_c$ disappears. These substrate-induced changes are more pronounced if the two spheres are close to the substrate (Fig.~\ref{D2_X}). Without a substrate the CCF between spheres of opposite BCs is purely repulsive. In the presence of a substrate the corresponding lateral CCF for $(-,+,+)$ BCs turns attractive for large lateral distances $L$ (Fig.~\ref{L_X}), which is a pure many-body effect.

We have also studied the direction of the total CCF acting on colloid (2) for various spatial configurations and BCs in order to assess the preferred arrangement of the colloids. For $(+,+,+)$ BCs we have found that they tend to aggregate laterally. In this case a collection of colloids with $(+)$ BCs, facing a substrate with the same BC, are expected to form a monolayer on the substrate. For the situation of $(-,-,+)$ BCs, we have found that the colloids are expected to aggregate on top of each other. This indicates that a set of several colloids with such BCs are expected to form three-dimensional clusters. These tendencies are enhanced upon approaching the critical point (Fig.~\ref{direction_resulting_force_D2}).

By calculating the pairwise colloid-colloid and colloid-substrate forces and subtracting them from the total force, we have determined the pure many-body contribution to the force acting on colloid (2). For the scaling functions associated with the normal many-body CCFs we have found the interesting feature of a change of sign at fixed temperature upon varying the lateral position of colloid (2) (Fig.~\ref{MB_L_Z}). This implies that, for a given temperature, there is a lateral position where the normal many-body CCF is zero, in which case the sum of pairwise forces provides a quantitatively reliable description of the interactions of the system. As expected we have found that the contribution of the many-body CCFs to the total force is large if the colloid-colloid and colloid-substrate distances are small, as well as if the binary liquid mixture is close to its critical point.

We have compared our results with corresponding ones for quantum-electrodynamic Casimir interactions. To this end we have referred to the results in Ref.~\cite{PhysRevA.83.042516} for two dielectric spheres immersed in ethanol and facing a plate. These authors analyze the influence of the distance $D=D_1=D_2$ from the plate on the equilibrium separation $L_D$ between the spheres, which are subject to quantum-electrodynamic Casimir forces. They find that the lateral many-body force is attractive (repulsive) if the stronger one of the two normal sphere-plate forces is attractive (repulsive) [Figs.~\ref{comparing}(a) and (b)]. On the other hand, in the case of CCFs we have found that for a configuration in which the surface-to-surface distance between the colloids is equal to the sphere-surface-to-substrate ones and equal to the radius of the spheres, the many-body contribution to the lateral CCF is always attractive, regardless of the BCs (Fig.~\ref{MB_X}).

We have also compared our results with the corresponding ones for the case of two atoms close to the planar surface of a solid body. In this respect we have referred to the McLachlan model~\cite{McLachlan1964} for the many-body contribution to the van der Waals potential [see Eq.~\eqref{outraequacaosuperimportantegente} and Fig.~\ref{mclachan_fig}] and the results from Ref.~\cite{PhysRevB.27.1314}. From this comparison we have found that if the two atoms are fixed at the same distance from the surface of the solid body, the normal many-body contribution to the total van der Waals force decays with the atom-atom distance $\rho$ as $\rho^{-8}$ for large atom-atom distances. This decay is much faster than the decay we estimate for the many-body contribution to the normal CCF, which within a suitable range appears to be slower than $L^{-2}$. Furthermore, we have found that the many-body contribution to the lateral van der Waals force is repulsive while the corresponding many-body CCF is attractive regardless of the set of BCs.

Finally we have considered the configuration in which the two colloids are vertically aligned with respect to the substrate (Fig.~\ref{vertical_sketch}). We have calculated the many-body contribution to the normal CCF acting on colloid (1) for two spherical colloids of the same size ($R_1=R_2=R$) keeping the sphere-surface-to-surface distance $S=R$ fixed (Fig.~\ref{MB_vertical_Z}). We have varied the sphere-surface-to-substrate distance $D_1$ for several BCs and have found that if the colloids have the same BCs,  the many-body contribution to the normal CCF is directed towards the substrate [Figs.~\ref{MB_vertical_Z}(a) and (d)], whereas for colloids with opposite BCs, the many-body contribution to the normal CCF is directed away from the substrate [Figs.~\ref{MB_vertical_Z}(b) and (c)]. We have found that the contribution of the many-body CCF to the total force is between 10$\%$ and 15$\%$.


\section*{Acknowledgments}

T.G.M. would like to thank S. Kondrat for valuable support with the computational tools used to perform the numerical calculations. S.D. thanks M. Cole for providing Ref.~\cite{PhysRevB.27.1314}.


\end{document}